\documentclass[aps,prc,preprint,superscriptaddress,showpacs]{revtex4}
\usepackage[dvips]{graphicx}
\usepackage{amsmath,amssymb,times}
\usepackage{ulem}

\newcommand{\beq}{\begin{equation}}
\newcommand{\eeq}{\end{equation}}
\newcommand{\bea}{\begin{eqnarray}}
\newcommand{\eea}{\end{eqnarray}}

\renewcommand{\a}{\alpha}

\DeclareSymbolFont{boldletters}{OML}{cmm} {b}{it}
\DeclareSymbolFontAlphabet{\mathbit}{boldletters}
\DeclareMathSymbol{\alpha}{\mathalpha}{letters}{"0B}
\DeclareMathSymbol{\beta}{\mathalpha}{letters}{"0C}
\DeclareMathSymbol{\gamma}{\mathalpha}{letters}{"0D}
\DeclareMathSymbol{\delta}{\mathalpha}{letters}{"0E}
\DeclareMathSymbol{\epsilon}{\mathalpha}{letters}{"0F}
\DeclareMathSymbol{\zeta}{\mathalpha}{letters}{"10}
\DeclareMathSymbol{\eta}{\mathalpha}{letters}{"11}
\DeclareMathSymbol{\theta}{\mathalpha}{letters}{"12}
\DeclareMathSymbol{\iota}{\mathalpha}{letters}{"13}
\DeclareMathSymbol{\kappa}{\mathalpha}{letters}{"14}
\DeclareMathSymbol{\lambda}{\mathalpha}{letters}{"15}
\DeclareMathSymbol{\mu}{\mathalpha}{letters}{"16}
\DeclareMathSymbol{\nu}{\mathalpha}{letters}{"17}
\DeclareMathSymbol{\xi}{\mathalpha}{letters}{"18}
\DeclareMathSymbol{\pi}{\mathalpha}{letters}{"19}
\DeclareMathSymbol{\rho}{\mathalpha}{letters}{"1A}
\DeclareMathSymbol{\sigma}{\mathalpha}{letters}{"1B}
\DeclareMathSymbol{\tau}{\mathalpha}{letters}{"1C}
\DeclareMathSymbol{\upsilon}{\mathalpha}{letters}{"1D}
\DeclareMathSymbol{\phi}{\mathalpha}{letters}{"1E}
\DeclareMathSymbol{\chi}{\mathalpha}{letters}{"1F}
\DeclareMathSymbol{\psi}{\mathalpha}{letters}{"20}
\DeclareMathSymbol{\omega}{\mathalpha}{letters}{"21}
\DeclareMathSymbol{\varepsilon}{\mathalpha}{letters}{"22}
\DeclareMathSymbol{\vartheta}{\mathalpha}{letters}{"23}
\DeclareMathSymbol{\varpi}{\mathalpha}{letters}{"24}
\DeclareMathSymbol{\varrho}{\mathalpha}{letters}{"25}
\DeclareMathSymbol{\varsigma}{\mathalpha}{letters}{"26}
\DeclareMathSymbol{\varphi}{\mathalpha}{letters}{"27}
\DeclareMathSymbol{\Gamma}{\mathalpha}{letters}{"00}
\DeclareMathSymbol{\Delta}{\mathalpha}{letters}{"01}
\DeclareMathSymbol{\Theta}{\mathalpha}{letters}{"02}
\DeclareMathSymbol{\Lambda}{\mathalpha}{letters}{"03}
\DeclareMathSymbol{\Xi}{\mathalpha}{letters}{"04}
\DeclareMathSymbol{\Pi}{\mathalpha}{letters}{"05}
\DeclareMathSymbol{\Sigma}{\mathalpha}{letters}{"06}
\DeclareMathSymbol{\Upsilon}{\mathalpha}{letters}{"07}
\DeclareMathSymbol{\Phi}{\mathalpha}{letters}{"08}
\DeclareMathSymbol{\Psi}{\mathalpha}{letters}{"09}
\DeclareMathSymbol{\Omega}{\mathalpha}{letters}{"0A}

\begin{document}
\preprint{SAGA-HE-243-08}
\title{Correlations among discontinuities in QCD phase diagram}

\author{Kouji Kashiwa}
\email[]{kashiwa@phys.kyushu-u.ac.jp}
\affiliation{Department of Physics, Graduate School of Sciences, Kyushu University,
             Fukuoka 812-8581, Japan}

\author{Yuji Sakai}
\email[]{sakai@phys.kyushu-u.ac.jp}
\affiliation{Department of Physics, Graduate School of Sciences, Kyushu University,
             Fukuoka 812-8581, Japan}

\author{Hiroaki Kouno}
\email[]{kounoh@cc.saga-u.ac.jp}
\affiliation{Department of Physics, Saga University,
             Saga 840-8502, Japan}

\author{Masayuki Matsuzaki}
\email[]{matsuza@fukuoka-edu.ac.jp}
\affiliation{Department of Physics, Fukuoka University of Education, 
             Munakata, Fukuoka 811-4192, Japan}

\author{Masanobu Yahiro}
\email[]{yahiro@phys.kyushu-u.ac.jp}
\affiliation{Department of Physics, Graduate School of Sciences, Kyushu University,
             Fukuoka 812-8581, Japan}

\date{\today}

\begin{abstract}
We show, in general,  that when a discontinuity of either 
zeroth-order or first-order 
takes place in an order parameter such as the chiral condensate, 
discontinuities of the same order emerge in other order parameters 
such as the Polyakov loop. 
A condition for the coexistence theorem to 
be valid is clarified. 
Consequently, only when the condition breaks down, 
zeroth-order and first-order discontinuities can coexist on a phase boundary. 
We show with 
the Polyakov-loop extended Nambu--Jona-Lasinio model that 
such a type of coexistence is realized 
in the imaginary chemical potential region of the QCD phase diagram. 
We also present examples of coexistence of the same-order discontinuities 
in the real chemical potential region. 
\end{abstract}

\maketitle

Exploring the phase diagram of Quantum Chromodynamics
(QCD) is one of the most important subjects in hadron physics. 
Actually, many works were done so far on this subject, and  
it is expected  that there appear several interesting phases 
in hot and/or dense quark matter; 
for example, chiral symmetry broken and restored phases, 
confinement and deconfinement phases, 
two-flavor color superconducting and color-flavor locked phases, 
and so on; for example, see Ref.~\cite{Alford} and references therein. 
These phases are characterized in terms of 
some exact or approximate order parameters such as 
the chiral condensate, the diquark condensate, the Polyakov loop, and so on. 
Therefore, correlations among these order parameters are to be investigated. 
In particular, the relation between orders of their discontinuities 
is essential. It was proven by 
Barducci, Casalbuoni, Pettini and Gatto (BCPG) \cite{BCGG} 
that different first-order phase transitions take place simultaneously. 
The theorem corresponds to a generalization of the Clausius-Clapeyron relation.

Studying these correlations directly in QCD is desired, 
however, in the finite chemical potential region, 
lattice QCD is still far from perfection because of the sign problem; 
for example, see Ref.~\cite{Kogut} and references therein.
So the phase diagram was investigated with effective models. 
Recently, important progress was made by 
the Polyakov-loop extended Nambu--Jona-Lasinio (PNJL) model
~\cite{Meisinger,Dumitru,Fukushima1,Fukushima2,Ghos,Megias,Ratti1,Ciminale,Ratti2,Rossner,Hansen,Sasaki,Schaefer,Costa,Kashiwa1,Fu,Abuki1,Abuki2,Sakai,Kouno}.
This model can describe the chiral, the color superconducting and 
the confinement/deconfinement phase transitions.

\begin{figure}[htbp]
\begin{center}
 \includegraphics[width=0.4\textwidth]{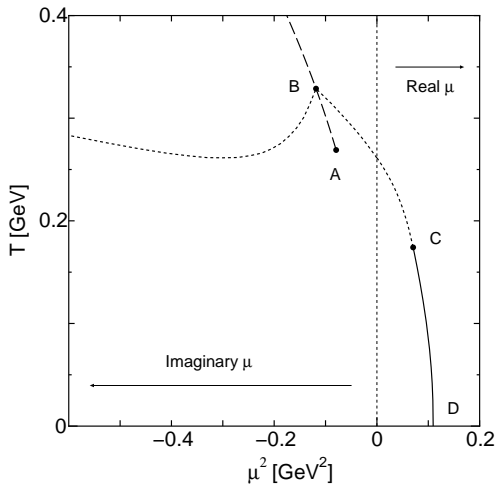} 
\end{center}
\caption{Phase diagram 
in the $\mu^2$-$T$ plane predicted by the PNJL model 
in the chiral limit.}
\label{PhaseDiagram-RI}
\end{figure}

Figure \ref{PhaseDiagram-RI} shows the phase diagram 
in the chiral limit predicted by the two-flavor PNJL model 
without diquark condensate; 
the details of the calculation 
will be shown latter.  The diagram is drawn in the $\mu^2$-$T$ plane, 
where $T$ stands for temperature and $\mu$ 
for quark chemical potential. 
The solid and dotted curves represent 
first- and second-order chiral phase transitions,  
respectively. 
In this paper, 
when an order parameter has a discontinuity in its value (zeroth-order), 
we call it the first-order phase transition. 
Meanwhile, when an order parameter has a discontinuity in its derivative 
(first order) and its susceptibility is divergent, we refer to it 
as the second-order phase transition. 
However, our discussion is mainly concentrated on the 
relationship between zeroth- and first-order discontinuities.

On the solid curve between points C and D, 
two zeroth-order discontinuities emerge simultaneously in 
the Polyakov loop and the chiral condensate. 
This is a typical example of the BCPG theorem. 
As an interesting fact, on the dotted curves, 
two first-order discontinuities take place 
simultaneously in the chiral condensate and the Polyakov loop. 
This implies 
the BCPG theorem on the zeroth-order discontinuity of order parameter 
can be extended to the case of the 
first-order one. 
As another interesting point, 
on the dashed curve between points A and B, 
a first-order discontinuity of the chiral condensate coexists 
with zeroth-order discontinuities of quark number density
and other $\theta$-odd quantities, where $\theta=-i\mu/T$. 
On the dashed curve moving up 
from point B, furthermore, the quark number density 
still has a zeroth-order discontinuity, 
although the chiral condensate is always zero. 
Thus, the relation between orders of 
discontinuities of various quantities is much richer 
than the BCPG theorem predicts.

In the left half plane of Fig. \ref{PhaseDiagram-RI}, 
$\mu$ is imaginary. 
However, the phase diagram in the region 
is also important, since in the region lattice QCD has no sign problem and 
then its results are available. 
Hence, the validity of the PNJL model can be tested there by comparing 
the model results with the lattice ones. 
Actually, it has been shown for the case of finite quark mass
that the results of the PNJL model are consistent with those of the lattice 
simulations \cite{Sakai}. 
Furthermore, the real $\mu$ system can be regarded as 
an image of the imaginary $\mu$ one, 
since the canonical partition function of real $\mu$ is 
the Fourier transform of 
the grand canonical partition function of imaginary $\mu$ \cite{RW}. 

The aim of this paper is to extend the BCPG theorem on 
the zeroth-order discontinuity of order parameter to the case of 
the first-order discontinuity, 
that is, we show that once a discontinuity of either zeroth-order 
or first-order 
takes place in an order parameter 
such as the chiral condensate, 
discontinuities of the same order appear in other 
order parameters such as the Polyakov loop. 
The original coexistence theorem of BCPG 
on the zeroth-order discontinuity and the present 
coexistence theorem on the first-order discontinuity are preserved, 
when the phase boundary is shifted in both the 
$T$ and $\mu$ directions by varying values of external parameters such 
as the current quark mass; 
the condition will be shown later in 
(\ref{condition-T}) and (\ref{condition-mu}). 
In other words, 
discontinuities in mutually different orders can coexist 
only when the condition breaks. 
Such a situation is not just a trivial exception 
but a physical relevance. 
Actually, we will show that the situation is realized in the 
the Roberge and Weiss (RW) phase transition \cite{RW} 
appearing in the imaginary chemical potential region of 
the QCD phase diagram, and prove from the viewpoint of 
the coexistence theorem that the RW phase transition is 
a family of zeroth- and first-order discontinuities. 
This resolution of the RW phase transition is a principal subject 
of the present paper. 
We present some examples of the coexistence by using the PNJL model 
for both real and imaginary chemical potential regions in the phase diagram. 

We begin with the grand canonical partition function 
\begin{align}
Z(T,\mu)&= {\rm Tr} \exp[-\beta({\hat H}-\mu {\hat N})]
\end{align} 
with a Hamiltonian of the form 
\bea
{\hat H}={\hat H}_0 + \sum_\alpha \lambda_\alpha {\hat {\cal O}_\alpha}, 
\eea
where ${\hat H_0}$ determines the intrinsic system,  
$\lambda_{\alpha}$ are external parameters conjugate to 
the hermitian operators ${\hat {\cal O}_\alpha}$ and 
$\beta=1/T$, 
$\mu$ is the chemical potential 
and ${\hat N}$ is the particle number. 
The thermodynamical potential $\Omega (T,\mu)$ is given by 
\begin{align}
\Omega (T,\mu) &=  -\frac{T}{V} \ln Z(T,\mu) 
\end{align}
with $V$ the three-dimensional volume, 
and the entropy density $s$ and the particle number density $n$ are also by  
\begin{align}
s = -\Bigl( \frac{\partial \Omega}{\partial T} \Bigr)_{\mu,\lambda},~~~~
n = -\Bigl( \frac{\partial \Omega}{\partial \mu} \Bigr)_{T,\lambda},
\end{align}
where the subscript $x$ means 
that $x$ is fixed in the partial differentiation. 
The expectation value of the operator 
${\hat {\cal O}_{\alpha}}$ per volume 
\begin{align}
o_{\alpha}=\frac{\langle {\hat {\cal O}}_{\alpha} \rangle}{V}
= \frac{1}{V Z} {\rm Tr}\{ {\hat {\cal O}_{\alpha}} 
\exp[-\beta({\hat H}-\mu{\hat N})]\} 
\end{align}
is given by 
\bea
o_{\alpha}=\Bigl( \frac{\partial \Omega}{\partial \lambda_\alpha} \Bigr)_{T,\mu,\lambda'} ,
\eea
where the subscript $\lambda'$ shows that all the $\lambda$ except $\lambda_\alpha$ are 
fixed in the partial differentiation. 
The subscripts of the partial differentiation will be suppressed 
for simplicity, unless any confusion arises.

\begin{figure}[htbp]
\begin{center}
 \includegraphics[width=0.4\textwidth]{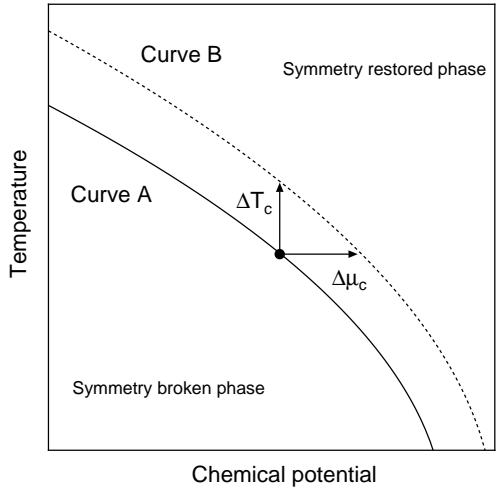} 
\end{center}
\caption{External parameter dependence of the phase boundary.
        Phase boundaries are projected on the $\mu$-$T$ plane 
        from the $\lambda_\alpha$-$\mu$-$T$ space.}
\label{Line}
\end{figure}

First we recapitulate the original BCPG theorem \cite{BCGG} 
on the zeroth-order discontinuity (the first-order phase transition) 
in order to know what is assumed in the proof. 
The proof is made as follows. 
We start with the assumption that 
there appears a discontinuity in $o_{\gamma}$, and show that 
the discontinuity propagates to other order parameters $o_{\alpha' \neq \gamma }$. 
Hereafter, $\alpha'$ stands for $\alpha$ except $\gamma$. 
Thus, no assumption is made beforehand on the property of 
discontinuities appearing in $o_{\alpha'}$. 
The first-order phase transition appearing in $o_{\gamma}$
is drawn 
by the solid curve $(\mu_{\rm c},T_{\rm c})$ schematically in Fig.~\ref{Line}; 
its typical example is the chiral transition at low temperature shown in 
Fig.~\ref{PhaseDiagram-RI}. 
The phase boundary (curve A) is shifted to curve B by taking different sets of 
external parameters, $\{\lambda_\alpha\}_B$.  
The thermodynamical potentials $\Omega_i$ of phases $i=1$ and 2 
on curve A satisfy the Gibbs condition 
\begin{align}
&\Omega_1(T_{\rm c}(\{\lambda_\alpha\}),\mu_{\rm c}(\{\lambda_\alpha\}),\{\lambda_\alpha\})
\nonumber \\
&=\Omega_2(T_{\rm c}(\{\lambda_\alpha\}),\mu_{\rm c}(\{\lambda_\alpha\}),\{\lambda_\alpha\}).
\label{Gibbs} 
\end{align}

Differentiating the thermodynamical potentials with respect to 
$\lambda_\gamma$ on the curve 
leads to 
\begin{align}
& \frac{\partial \Omega_1}{\partial \lambda_\gamma} 
\Bigl|_{\rm c}
+\frac{\partial \Omega_1}{\partial T} \Bigl|_{\rm c} \frac{\partial T_{\rm c}}
{\partial \lambda_\gamma} 
+\frac{\partial \Omega_1}{\partial \mu} \Bigl|_{\rm c} \frac{\partial \mu_{\rm c}}
{\partial \lambda_\gamma} 
\nonumber \\
&=
 \frac{\partial \Omega_2}{\partial \lambda_\gamma} \Bigl|_{\rm c}
+\frac{\partial \Omega_2}{\partial T} \Bigl|_{\rm c} \frac{\partial T_{\rm c}}
{\partial \lambda_\gamma} 
+\frac{\partial \Omega_2}{\partial \mu} \Bigl|_{\rm c} \frac{\partial \mu_{\rm c}}
{\partial \lambda_\gamma} ,
\end{align}
where 
the subscript $|_{\rm c}$ denotes that the quantities are evaluated
at $(\mu_{\rm c},T_{\rm c})$.
Hence we obtain 
\begin{align}
\delta o_{\gamma}  
&= \frac{\partial T_{\rm c}}{\partial \lambda_\gamma} \delta s  
  +\frac{\partial \mu_{\rm c}}{\partial \lambda_\gamma} \delta n ,
\label{g-CC}
\end{align}
where $\delta x=x_{1}-x_{2}$ ($x=o_{\gamma}, s, n$) is evaluated on 
the phase boundary. 
In Fig. \ref{Line}, 
the correspondence between each individual point on curve A and that 
on curve B 
is not unique. This means that one can define an infinitesimal 
variation of $T_{\rm c}$ in the $T$ direction with fixed $\mu_{\rm c}$, 
\bea
T_{\rm c}(\lambda_\gamma + \Delta \lambda_\gamma) - T_{\rm c}(\lambda_\gamma)
=\frac{\partial T_{\rm c}}{\partial \lambda_\gamma} \Bigl|_{\mu_{\rm c}} 
\Delta \lambda_\gamma , 
\eea
and an infinitesimal variation $\mu_{\rm c}$ in the $\mu$ direction with 
fixed $T_{\rm c}$,
\bea
\mu_{\rm c}(\lambda_\gamma + \Delta \lambda_\gamma) - \mu_{\rm c}(\lambda_\gamma)
=\frac{\partial \mu_{\rm c}}{\partial \lambda_\gamma} \Bigl|_{T_{\rm c}} 
\Delta \lambda_\gamma , 
\eea
where the subscript $|_x$ denotes that $x$ is fixed; 
these variations are illustrated by the arrows in Fig. \ref{Line}. 
Using these variations, one can see from (\ref{g-CC}) that 
\bea
\delta o_{\gamma} 
= \frac{\partial T_{\rm c}}{\partial \lambda_\gamma} \Bigl|_{\mu_{\rm c}} \delta s  
= \frac{\partial \mu_{\rm c}}{\partial \lambda_\gamma} \Bigl|_{T_{\rm c}} \delta n . 
\label{general-CC}
\eea
We find from $\delta o_{\gamma} \neq 0$ that 
${\partial T_{\rm c}}/{\partial \lambda_\gamma}|_{\mu_c}$ and 
${\partial \mu_{\rm c}}/{\partial \lambda_\gamma}|_{T_c}$ are nonzero, 
since $\delta s$ and $\delta n$ never diverge.

When all the $\lambda_{\a}$'s are fixed at zero, 
$\mu_{\rm c}$ can be regarded as a function of $T_{\rm c}$: 
$\mu_{\rm c}=\mu_{\rm c}(T_{\rm c})$.
Differentiating \eqref{Gibbs} with respect to 
$T_{\rm c}$, one can get 
\begin{align}
\frac{dT_{\rm c}}{d\mu_{\rm c}}&=-\frac{\delta n}{\delta s}. 
\label{CC}
\end{align}
Equation~\eqref{CC} is the Clausius-Clapeyron relation 
between $\delta s$ and $\delta n$~\cite{Halasz}, and 
Eq.~\eqref{general-CC} is a generalization of the relation 
to the case of nonzero $\lambda_{\a}$.

A relation similar to \eqref{general-CC} is obtainable 
for $\alpha'$:
\bea
\delta o_{\alpha'} 
= \frac{\partial T_{\rm c}}{\partial \lambda_{\alpha'}} 
\Bigl|_{\mu_{\rm c}} \delta s  
= \frac{\partial \mu_{\rm c}}{\partial \lambda_{\alpha'}} 
\Bigl|_{T_{\rm c}} \delta n .
\eea
Here it should be noted that 
the curve $(\mu_{\rm c},T_{\rm c})$ is defined by a discontinuity 
appearing in $o_{\gamma}$. 
The discontinuity $\delta o_{\gamma}\neq 0$ induces a 
new discontinuity $\delta o_{\alpha'}\neq 0$ 
through $\delta s\neq 0$, when
\bea
\frac{\partial T_{\rm c}}{\partial \lambda_{\alpha'}} \Bigl|_{\mu_{\rm c}}\neq 0 . 
\label{condition-T}
\eea
Similarly, the discontinuity $\delta o_{\gamma}\neq 0$ induces 
$\delta o_{\alpha'}\neq 0$ through $\delta n\neq 0$, when
\bea
\frac{\partial \mu_{\rm c}}{\partial \lambda_{\alpha'}} \Bigl|_{T_{\rm c}} \neq 0. 
\label{condition-mu}
\eea
Thus, when the conditions (\ref{condition-T}) and (\ref{condition-mu}) 
are satisfied, two first-order phase transitions take place simultaneously. 
In other words, the discontinuity of $o_\gamma$ 
propagates to other physical quantity $o_{\alpha'}$ through those of $s$ 
and $n$. 
The conditions mean 
that the phase boundary is shifted in both the $T$ and 
$\mu$ directions in the $\mu$-$T$ plane by varying $\lambda_{\alpha'}$.

An early application of the BCPG theorem was to the case of a 2+1 flavor 
model in which two chiral condensations 
exist~\cite{Tawfik}. 
A similar situation is expected when an isospin chemical potential is 
introduced in 2 flavor models if a flavor mixing interaction is 
included~\cite{Frank}.

Here we show an example of the simultaneous occurrence of 
zeroth-order discontinuities of order parameters 
by using the PNJL model in the chiral limit. 
The formulation and the parameter set of the PNJL model
are given in Refs.~\cite{Kashiwa1,Kashiwa2}, 
where the pure gauge part is obtained
by reproducing lattice QCD data in
the pure gauge theory~\cite{Boyd,Kaczmarek}
as shown in Ref.~\cite{Ratti1}.
In the present paper, we put $m_0=0$ with keeping other parameters unchanged.

In the chiral limit, the chiral condensate 
$\sigma=\langle {\bar q} q \rangle$ is an exact order 
parameter of the spontaneous chiral symmetry breaking:
namely, $o_{\gamma}=\sigma$ and $\lambda_{\gamma}=m_0$ for 
the quark field $q$ and the current quark mass $m_0$. 
The Polyakov loop $\Phi$ is an exact order parameter of the spontaneous 
${\mathbb Z}_3$ symmetry breaking in the pure gauge theory, but 
the symmetry is not exact anymore 
in the system with dynamical quarks. However, 
$\Phi$ 
still seems to be a good indicator of the deconfinement phase transition. 
We then regard $\Phi$ as an approximate order parameter of 
the deconfinement phase transition.

\begin{figure}[htbp]
\begin{center}
 \includegraphics[width=0.4\textwidth]{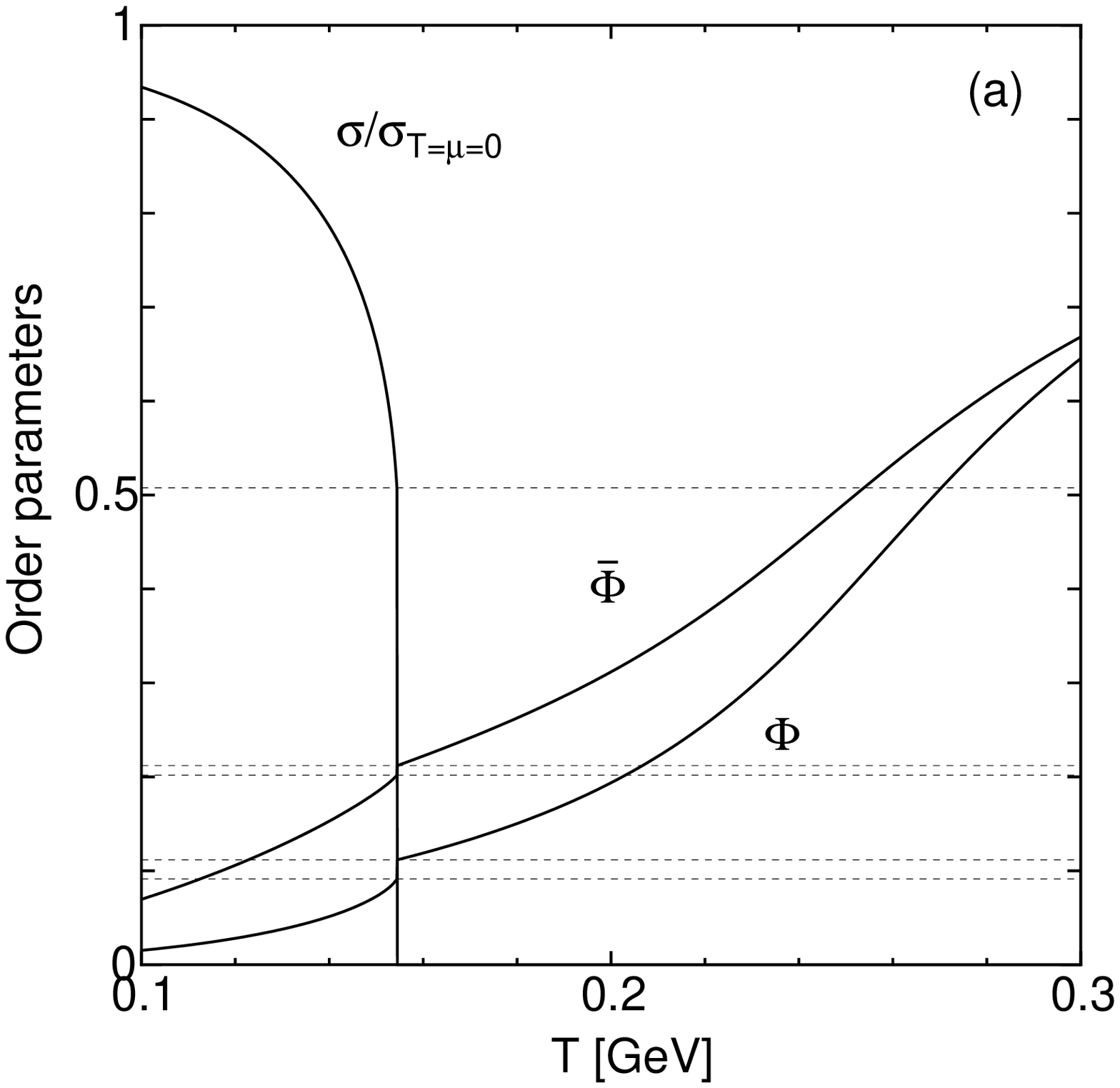} 
 \includegraphics[width=0.4\textwidth]{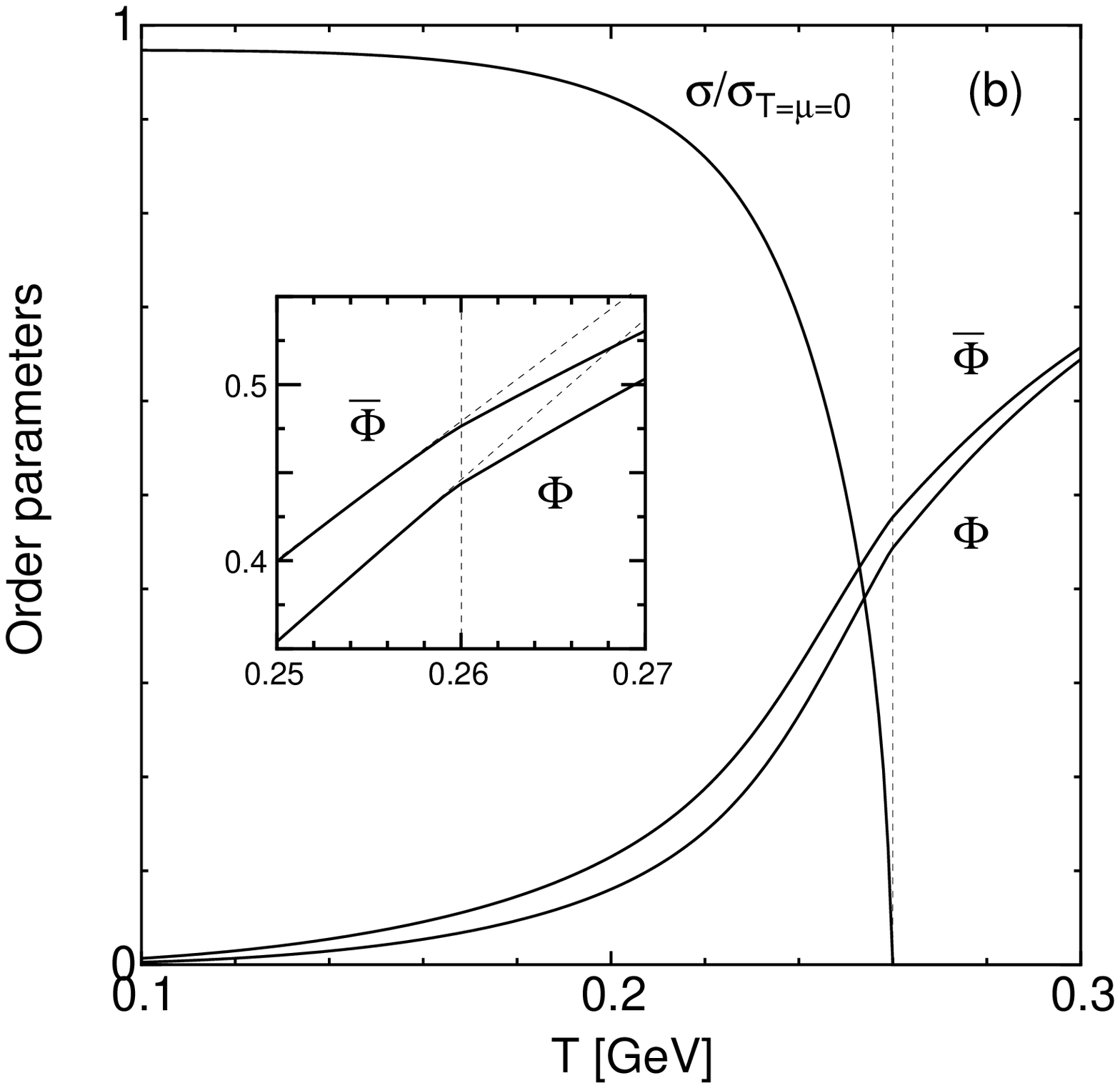} 
\end{center}
\caption{The temperature dependence
of the chiral condensate $\sigma$, the Polyakov loop 
$\Phi$ and its conjugate ${\bar \Phi}$ at (a) $\mu=280$ MeV and (b) $\mu=50$ MeV 
in the chiral limit. 
The chiral condensate is normalized by the value at $T=\mu=0$. 
The inset figure in (b) represents $\Phi$ and ${\bar \Phi}$ near $T=260$~MeV 
that is the critical temperature.}
\label{PNJL-12}
\end{figure}

Figure~\ref{PNJL-12}(a) represents $T$ dependence of 
$\sigma$, $\Phi$ and the charge-conjugated Polyakov loop 
${\bar \Phi}$ at $\mu=280$ MeV. These are 
discontinuous at the same temperature $T=154$ MeV.
This behavior is consistent with the BCPG theorem that 
guarantees the simultaneous occurrence of zeroth-order discontinuities 
of order parameters. Here one can also find that 
$\mu$ dependence of $\Phi$ is similar to that of ${\bar \Phi}$. 
This is true for other real $\mu$.

The BCPG theorem does not necessarily mean that 
there can not exist a quarkyonic phase~\cite{McLerran1} 
defined in the limit of large number of colors as a phase that 
has a finite value of the baryon number density $n$ but is confined. 
This is understandable as follows. 
The quarkyonic phase was recently investigated with 
the PNJL model~\cite{Fukushima2,Abuki2,McLerran2} and 
the strong coupling QCD~\cite{Miura}. The PNJL analysis of 
Ref.~\cite{Abuki2} shows the simultaneous occurrence of 
zeroth-order discontinuities of $\sigma$, $\Phi$ and $n$. 
However, the discontinuity in $\Phi$ is only a jump from 
a small value to another small one, 
while that in $n$ is a jump from almost 0 to a value larger than 
the nuclear saturation density. 
As mentioned above, $\Phi$ is only an approximate order parameter of 
the deconfinement phase transition, and then such a small jump is possible. 
Such a small jump could be just a propagation of the discontinuity 
in $\sigma$. 
Thus, we can not say necessarily from 
the small jump of $\Phi$ that 
a first-order deconfinement phase takes place together with 
the chiral phase transition and the phase transition of $n$. 
A plausible definition of a critical temperature of 
the deconfinement phase transition is a temperature that gives 
$\Phi=0.5$. In this definition, 
the deconfinement transition is crossover, and 
then the BCPG theorem is not applicable anymore.

Next, we proceed to the case that an order parameter has 
a first-order discontinuity. In this case, 
since  a first-order discontinuity 
becomes continuous by a change of external parameters, 
the boundary must be defined in terms of a 
susceptibility as follows. 
Here we take the chiral transition in 2 flavor systems 
at high temperature, shown by the dotted curve in Fig.~\ref{PhaseDiagram-RI}, 
as a typical example: 
namely, $\lambda_\gamma=m_0$ and 
$o_\gamma=\sigma=\langle \bar{q}q \rangle$. 
The second-order chiral phase transition at $m_0=0$ 
becomes crossover whenever $m_0$ is finite \cite{Sakai}. 
It is possible to define the phase boundary of such a crossover 
with the chiral susceptibility 
$\chi_\sigma=- \partial \sigma/\partial m_0$ 
so that the $T$ dependence of $\chi$ becomes maximum on the boundary. 
This definition works also in the chiral limit, although 
the maximum is infinity. 
In this case, curves A and B in Fig. \ref{Line} are 
reinterpreted as phase boundaries so defined and correspond to 
the chiral limit ($m_0=0$) and to the case of 
small $m_0$, respectively. Curve A can move continuously and 
reach curve B by varying $m_0$ from 0 to a finite value. 
The chiral susceptibility $\chi_\sigma$ is divergent 
on the boundary A, since the chiral phase transition is of second order there.

Now, we consider curve A defined above and its vicinity. 
The system concerned has no zeroth-order discontinuity, 
$\delta s=\delta n=0$. 
Differentiating $\delta s=0$ 
with respect to $\lambda_\gamma$ on the boundary $(\mu_{\rm c},T_{\rm c})$ leads to
\bea
\delta \Bigl( \frac{\partial s}{\partial \lambda_\gamma} \Bigr) 
+\frac{\partial T_{\rm c}}{\partial \lambda_\gamma}
 \delta \Bigl( \frac{\partial s}{\partial T}  \Bigr)  
+\frac{\partial \mu_{\rm c}}{\partial \lambda_\gamma}
 \delta \Bigl( \frac{\partial s}{\partial \mu}  \Bigr)=0. 
\eea
Using the relation 
${\partial s}/{\partial \lambda_\gamma}=-{\partial o_\alpha}/{\partial T}$
and the variations in the $T$ and $\mu$ directions mentioned above, 
one can get 
\bea
\delta \Bigl( \frac{\partial o_\gamma}{\partial T} \Bigr)
=\frac{\partial T_{\rm c}}{\partial \lambda_\gamma} \Bigl|_{\mu_{\rm c}}
 \delta \Bigl( \frac{\partial s}{\partial T}  \Bigr) 
=\frac{\partial \mu_{\rm c}}{\partial \lambda_\gamma}\Bigl|_{T_{\rm c}}
 \delta \Bigl( \frac{\partial s}{\partial \mu}  \Bigr). 
\label{2nd-propagation-1}
\eea
Taking the same procedure for $\delta n=0$, one also obtains 
\bea
\delta \Bigl( \frac{\partial o_\gamma}{\partial \mu} \Bigr)
=\frac{\partial T_{\rm c}}{\partial \lambda_\gamma} \Bigl|_{\mu_{\rm c}}
 \delta \Bigl( \frac{\partial n}{\partial T}  \Bigr) 
=\frac{\partial \mu_{\rm c}}{\partial \lambda_\gamma}\Bigl|_{T_{\rm c}}
 \delta \Bigl( \frac{\partial n}{\partial \mu}  \Bigr). 
\label{2nd-propagation-2}
\eea
Other order parameters $o_{\alpha'}$ satisfy the same equations as 
(\ref{2nd-propagation-1}) and (\ref{2nd-propagation-2}). 
Note that all the equations are evaluated 
in the chiral limit $\lambda_\gamma=m_0=0$. 
It is found from (\ref{2nd-propagation-1}) and (\ref{2nd-propagation-2}) 
for $o_{\gamma}$ and the corresponding equations for $o_{\alpha'}$
that discontinuities 
$\delta ( {\partial o_\gamma}/{\partial T}) \neq 0 $ and 
$\delta ( {\partial o_\gamma}/{\partial \mu}) \neq 0 $ 
induce new ones 
$\delta ( {\partial o_{\alpha'}}/{\partial T}) \neq 0$ and 
$\delta ( {\partial o_{\alpha'}}/{\partial \mu}) \neq 0 $, 
when the conditions (\ref{condition-T}) and (\ref{condition-mu}) are 
satisfied. 
Thus, two first-order discontinuities of order parameters can coexist under 
the conditions (\ref{condition-T}) and (\ref{condition-mu}). 
Furthermore, it is found from 
(\ref{2nd-propagation-1}) and (\ref{2nd-propagation-2}) that 
$\delta ({\partial o_\gamma}/{\partial \mu})$ is not zero 
whenever $\delta ({\partial o_\gamma}/{\partial T})$ is not zero, 
because of $\partial n / \partial T =\partial s / \partial \mu$. 
Accordingly the first-order discontinuity of $o_\gamma$ emerges in both 
${\partial o_\gamma}/{\partial T}$ and ${\partial o_\gamma}/{\partial \mu}$. 

Here we show an example of the simultaneous occurrence of two discontinuities 
by the PNJL model in the chiral limit. 
Figure~\ref{PNJL-12}(b) represents $T$ dependence of 
$\sigma$, $\Phi$ and ${\bar \Phi}$ at $\mu=50$ MeV. 
Obviously, $\Phi$ and ${\bar \Phi}$ are not smooth at $T_{\rm c}=260$~MeV.
In the inset figure of Fig.~\ref{PNJL-12}(b), 
the solid curves show $\Phi$ and ${\bar \Phi}$ near $T_{\rm c}=260$~MeV, 
and two dotted lines do tangential lines of the solid curves 
at $T=T_{\rm c} - 0$. 
The deviations between the solid curves and the corresponding dotted lines 
indicate that 
$\Phi$ and ${\bar \Phi}$ are not smooth at $T_{\rm c}=260$~MeV.
Thus, $\partial \sigma/\partial T$, $\partial \Phi/\partial T$ 
and $\partial {\bar \Phi}/\partial T$
are discontinuous at the same temperature, as expected from 
the coexistence theorem on the first-order discontinuity of order parameter.

\begin{figure}[htbp]
\begin{center}
 \includegraphics[width=0.4\textwidth]{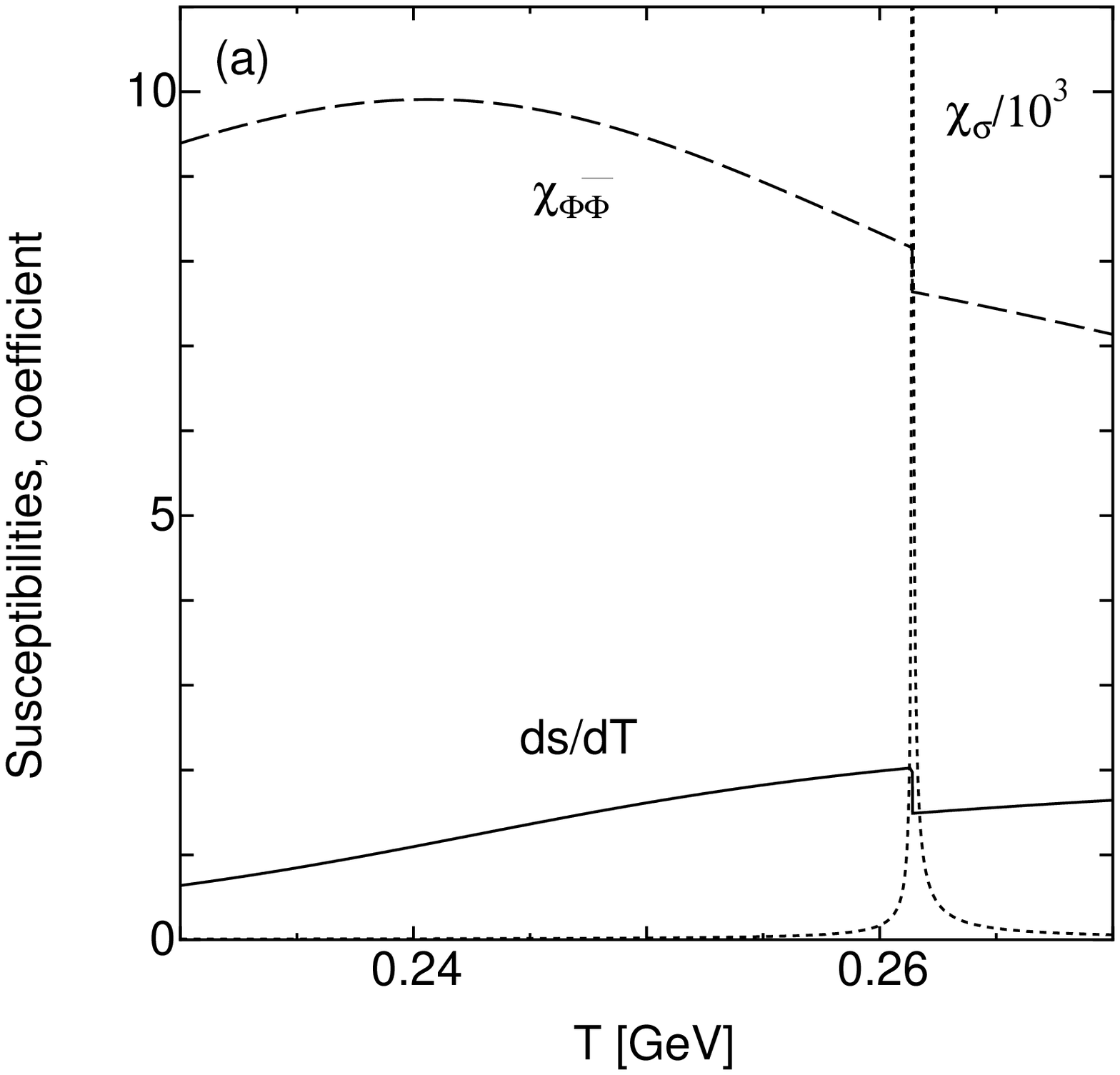} 
 \includegraphics[width=0.4\textwidth]{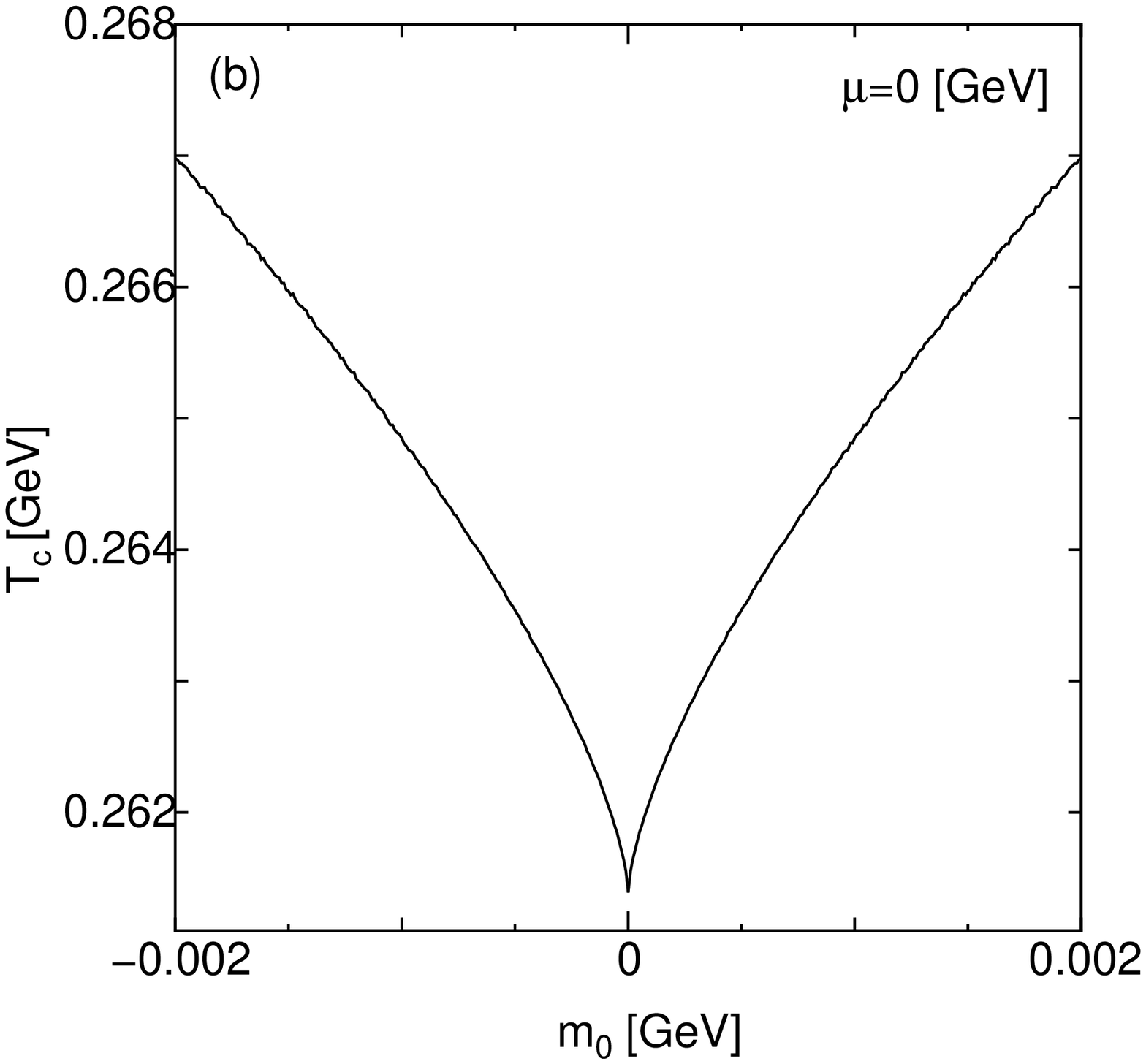} 
\end{center}
\caption{ Properties of thermal quantities at $\mu=0$:
(a) $T$ dependence of chiral and Polyakov-loop susceptibilities and 
$\partial s/\partial T$ at $m_0=0$ and (b) quark-mass dependence 
of the critical temperature. 
In panel (a), the solid, dashed and dotted curves represent 
$\partial s/\partial T$, $\chi_{\Phi{\bar \Phi}}$ and $\chi_{\sigma}$, 
respectively. The chiral susceptibility $\chi_{\sigma}$ has a divergent peak 
at $T=T_{\rm c}=261.4$~MeV. 
}
\label{m0-Tc}
\end{figure}

As shown in Fig.~\ref{PNJL-12}(b), 
$\delta(\partial \sigma/\partial T)$ diverges at $T=T_{\rm c}$, 
because  $\partial \sigma/\partial T|_{T=T_c-0}=\infty $ and 
$\partial \sigma/\partial T|_{T=T_{\rm c}+0}=0$; note that $\sigma \le 0$. 
Similar divergence is also seen 
on the second-order chiral phase transition line (dotted curves) 
in Fig.~\ref{PhaseDiagram-RI}.
This divergence indicates from \eqref{2nd-propagation-1} that 
$\partial T_{\rm c}/\partial m_0|_{\mu_{\rm c}}$ and/or 
$\delta ( {\partial s}/{\partial T})$ diverges 
on the second-order chiral phase transition curve. 
If $\delta ( {\partial s}/{\partial T})$ is infinite there, 
the divergence will propagate to other quantities 
$\delta ( {\partial o_{\a'}}/{\partial T})$ when 
the condition (\ref{condition-T}) is satisfied. 
As shown below, this is not 
the case of the second-order chiral phase transition. 
Figure~\ref{m0-Tc} presents 
$T$ dependence of $\partial s/\partial T$, 
the chiral susceptibility $\chi_\sigma$ and the Polyakov-loop susceptibility 
$\chi_{\Phi{\bar \Phi}}$ at $\mu=m_0=0$ in panel (a) and 
$m_0$ dependence of $T_c$ at $\mu=0$ in panel (b); 
definitions of $\chi_\sigma$ and $\chi_{\Phi{\bar \Phi}}$ are shown below. 
As shown in panel (a), $\chi_\sigma$ (the dotted curve) 
diverges at $T=261.4$~MeV, but 
$\partial s/\partial T$ (the solid curve) has a finite gap there. 
Meanwhile, panel (b) shows that 
the gradient $\partial T_{\rm c}/\partial m_0$ is divergent at $m_0=0$. 
In the present case, thus, 
the divergence in $\delta(\partial \sigma/\partial T)$ does not 
propagate to other quantities $\delta(\partial o_{\a'}/\partial T)$. 
Accordingly, the coexistence of first-order discontinuities of 
order parameters takes place, but 
the coexistence of second-order phase transitions does 
not occur necessarily, because there is a possibility that 
$\partial T_{\rm c}/\partial \lambda_{\gamma}|_{\mu_{\rm c}}$ and 
$\partial \mu_{\rm c}/\partial \lambda_{\gamma}|_{T_{\rm c}}$
diverge. 
In other words, when $\partial T_{\rm c}/\partial \lambda_{\gamma}|_{\mu_{\rm c}}$ 
and $\partial T_{\rm c}/\partial \lambda_{\a'}|_{\mu_{\rm c}}$
are nonzero and finite, 
or 
when  
$\partial \mu_{\rm c}/\partial \lambda_{\gamma}|_{T_{\rm c}}$ 
and $\partial \mu_{\rm c}/\partial \lambda_{\a'}|_{T_{\rm c}}$ 
are nonzero and finite,
the coexistence of second-order phase transitions takes place.

Susceptibilities $\chi_{ij}$ of $\sigma$, $\Phi$ and ${\bar \Phi}$ 
can be written as~\cite{Fukushima1,Sasaki,Kashiwa1} 
\begin{eqnarray}
\chi_{ij} &=& (K^{-1})_{ij}
~~~~(i,j=\sigma, \Phi, {\bar \Phi}),
\end{eqnarray}
where  
\begin{eqnarray}
K = \left(
\begin{array}{cccc} 
\frac{\beta}{4G^2_\mathrm{s}\Lambda} \frac{\partial^2 \Omega}{\partial \sigma^2} 
& 
-\frac{\beta}{2G_\mathrm{s}\Lambda^2} \frac{\partial^2 \Omega}{\partial \sigma \partial \Phi} 
& 
-\frac{\beta}{2G_\mathrm{s}\Lambda^2}\frac{\partial^2 \Omega}{\partial \sigma \partial {\bar \Phi}} \\

-\frac{\beta}{2G_\mathrm{s}\Lambda^2}\frac{\partial^2 \Omega}{\partial \Phi \partial \sigma} 
& 
\frac{\beta}{\Lambda^3} \frac{\partial^2 \Omega}{\partial \Phi^2} 
& 
\frac{\beta}{\Lambda^3} \frac{\partial^2 \Omega}{\partial \Phi \partial {\bar \Phi}} \\
-\frac{\beta}{2G_\mathrm{s}\Lambda^2} \frac{\partial^2 \Omega}{\partial {\bar \Phi} \partial \sigma} 
& 
\frac{\beta}{\Lambda^3} \frac{\partial^2 \Omega}{\partial {\bar \Phi} \partial \Phi} 
& 
\frac{\beta}{\Lambda^3} \frac{\partial^2 \Omega}{\partial {\bar \Phi}^2} \\
\end{array}
\right) ,
\end{eqnarray}
is a symmetric matrix of curvatures of $\Omega$ and $(K^{-1})_{ij}$ is an 
$(i,j)$ element of the inverse matrix $K^{-1}$. 
In the chiral limit, $\Omega$ is invariant under 
the transformation $\sigma \to - \sigma$~\cite{Kashiwa1} and hence 
$\sigma$-even. In the case that the chiral phase transition is 
the second order, as shown in Fig.~\ref{PNJL-12}, 
$\Omega$ becomes minimum at $\sigma=0$ when $T \geq T_{\rm c}$. 
Therefore, $K_{\sigma \Phi}$ and $K_{\sigma {\bar \Phi}}$ are 
zero at $T \geq T_{\rm c}$, because they are $\sigma$-odd. 
In this situation, the $\chi_{ij}$ at $T \geq T_{\rm c}$ are reduced to 
\begin{eqnarray}
\chi_{\sigma} \equiv 
\chi_{\sigma\sigma}=\frac{1}{K_{\sigma\sigma}}, \quad 
\chi_{ij} = (K_2^{-1})_{ij}~~~(i,j=\Phi, {\bar \Phi}),
\label{sus-2}
\end{eqnarray}
where 
\begin{eqnarray}
K_2 = \left(
\begin{array}{cccc} 
\frac{\beta}{\Lambda^3} \frac{\partial^2 \Omega}{\partial \Phi^2} 
& 
\frac{\beta}{\Lambda^3} \frac{\partial^2 \Omega}{\partial \Phi \partial {\bar \Phi}} 
\\
\frac{\beta}{\Lambda^3} \frac{\partial^2 \Omega}{\partial {\bar \Phi} \partial \Phi} 
& 
\frac{\beta}{\Lambda^3} \frac{\partial^2 \Omega}{\partial {\bar \Phi}^2} 
\end{array}
\right).
\end{eqnarray}
Thus, the susceptibilities of $\Phi$ and ${\bar \Phi}$ are decoupled 
from that of $\sigma$. In particular at $T = T_{\rm c}$,  
the curvature $K_{\sigma\sigma}$ is zero and 
then $\chi_{\sigma}$ is divergent, while at $T > T_{\rm c}$ 
the curvature $K_{\sigma\sigma}$ is positive and then $\chi_{\sigma}$ 
is a positive finite value. 
This divergence makes no influence on other susceptibilities 
$\chi_{ij}~(i,j=\Phi, {\bar \Phi})$, 
since $K_{\sigma \Phi}$ and $K_{\sigma {\bar \Phi}}$ are 
zero; see Refs.~\cite{Fujii,Fujii2} for the detail. 
Figure~\ref{m0-Tc}(a) is a typical example of this situation; 
$\chi_{\sigma}$ has a divergent peak, 
while $\chi_{\Phi{\bar \Phi}}$ does not.

In the imaginary chemical potential region ($\mu^2<0$), as shown 
by the dashed curve between points A and B of Fig. \ref{PhaseDiagram-RI}, 
there coexists a zeroth-order discontinuity of quark number density $n$ and 
a first-order discontinuity of chiral condensate $\sigma$. 
The coexistence is 
consistent with the proofs mentioned above, as shown below. 
This is the principal subject of the present paper.

\begin{figure}[htbp]
\begin{center}
 \includegraphics[width=0.4\textwidth]{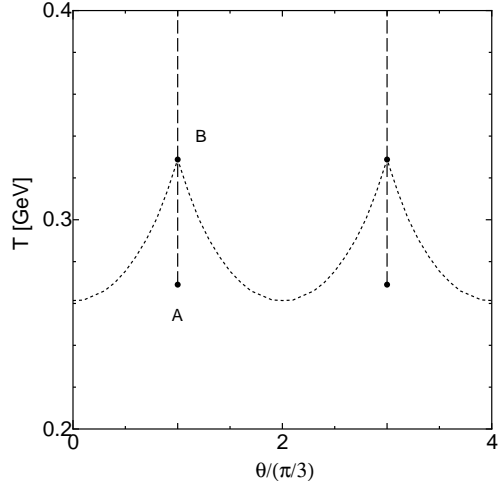} 
\end{center}
\caption{Phase diagram on the $\theta$-$T$ plane predicted by the PNJL model 
in the chiral limit.}
\label{PhaseDiagram-I}
\end{figure}

It is convenient to introduce a new variable $\theta=-i \mu/T$ instead of 
$\mu$. 
The conditions (\ref{condition-T}) and (\ref{condition-mu}) 
are then changed into 
\bea
\frac{\partial T_{\rm c}}{\partial \lambda_{\alpha'}} \Bigl|_{\theta_{\rm c}} 
\neq 0 , 
\\
\label{condition-T-2}
\frac{\partial \theta_{\rm c}}{\partial \lambda_{\alpha'}} \Bigl|_{T_{\rm c}} 
\neq 0 .
\label{condition-mu-2}
\eea
As shown later in Fig. \ref{PhaseDiagram-I}, 
the coexistence of $\delta n \neq 0$, $\delta \sigma = 0$ and 
$\delta (\partial \sigma /\partial \theta) \neq 0$ 
always appears on vertical lines $\theta=(2k+1)\pi/3$ 
in the $\theta$-$T$ plane, where $k$ is an integer. This indicates that 
${\partial \theta_{\rm c}}/{\partial m_0}=0$ and then 
the condition (\ref{condition-mu-2}) breaks down. 
Taking the new variable $\theta$ also changes (\ref{g-CC}) into 
\bea
\delta \tilde{\sigma}  
= \frac{\partial T_{\rm c}}{\partial m_0} \delta \tilde{s}  
  +\frac{\partial \theta_{\rm c}}{\partial m_0} \delta \tilde{n} 
= \frac{\partial T_{\rm c}}{\partial m_0} \delta \tilde{s} 
\label{g-CC-2}
\eea
with
\bea
&\tilde{s} = -\Bigl( \frac{\partial \Omega}{\partial T} \Bigr)_{\theta,\lambda}, 
\quad
\tilde{n} = -\Bigl( \frac{\partial \Omega}{\partial \theta} \Bigr)_{T,\lambda}
=i n T, \\
&\tilde{\sigma} =\Bigl( \frac{\partial {\Omega}}
{\partial m_0} \Bigr)_{\theta,T,\lambda'}
=\sigma ,
\eea
where use has been made of ${\partial \theta_{\rm c}}/{\partial m_0}=0$ 
in the second equality of (\ref{g-CC-2}). 
Thus, even if $\delta \tilde{n}$ is not zero, 
one can keep $\delta \sigma=0$ in (\ref{g-CC-2}) 
when $\delta \tilde{s}$ is zero.

The discontinuity $\delta \tilde{n} \neq 0$ was first pointed out 
by Roberge and Weiss (RW) \cite{RW}, and often called the RW transition.
Here, we consider how the discontinuity $\delta \tilde{n} \neq 0$ influences 
other order parameters $o_{\alpha}$. 
In this case, curve A in Fig. \ref{Line} is defined by the 
discontinuity $\delta \tilde{n} \neq 0$. 
The quantity $\delta \tilde{n}$ is a function of 
$T_{\rm c}$, $\theta_{\rm c}$ and $\lambda_\alpha$, 
but $\theta_{\rm c}$ does not depend on  $\lambda_\alpha$; namely, 
$\delta \tilde{n} = -f(T_{\rm c}(\{\lambda_\alpha \}),\theta_{\rm c},\{\lambda_\alpha \})$. 
Differentiating $\delta \tilde{n} + f=0$ with respect to $\lambda_\alpha$ leads to 
\bea
\delta \Bigl( \frac{\partial \tilde{n}}{\partial \lambda_\alpha} \Bigr)  
+\frac{\partial T_{\rm c}}{\partial \lambda_\alpha}
 \delta \Bigl( \frac{\partial \tilde{n}}{\partial T}  \Bigr)  
+ \frac{\partial f}{\partial \lambda_\alpha} \Bigl|_{\tilde{{\rm c}}}  
+ \frac{\partial f}{\partial T} \Bigl|_{\tilde{{\rm c}}}  
\frac{\partial T_{\rm c}}{\partial \lambda_\alpha} 
=0 
\eea
because of $\partial \theta_{\rm c}/\partial \lambda_\alpha =0$,
where the subscript $|_{\tilde{{\rm c}}}$ denotes that the quantities are evaluated 
at $(\theta_{\rm c}, T_{\rm c})$. 
Using $\partial \tilde{n} / \partial \lambda_\alpha 
=- \partial o_\alpha / \partial \theta$ and 
taking the variation in the $\theta$ direction with fixed $T_{\rm c}$, 
one can obtain 
\bea
\delta \Bigl( \frac{\partial o_\alpha}{\partial \theta} \Bigr) 
= \frac{\partial f}{\partial \lambda_\alpha} \Bigl|_{\tilde{{\rm c}}} .
\label{1st-2nd-propagation-1}
\eea
Taking the same procedure for $\delta \tilde{s}=0$ leads to 
\bea
\delta \Bigl( \frac{\partial o_\alpha}{\partial T} \Bigr)
=\frac{\partial \theta_{\rm c}}{\partial \lambda_\alpha}\Bigl|_{T_{\rm c}}
 \delta \Bigl( \frac{\partial \tilde{s}}{\partial \theta}  \Bigr) 
 \Bigl|_{\tilde{{\rm c}}}=0 
\label{1st-2nd-propagation-2}
\eea
because of ${\partial \theta_{\rm c}}/{\partial \lambda_\alpha}|_{T_{\rm c}}=0$.
Equations (\ref{1st-2nd-propagation-1}) and (\ref{1st-2nd-propagation-2}) 
indicate that order parameters 
are discontinuous in 
${\partial o_\alpha}/{\partial \theta}$ but not in 
${\partial o_\alpha}/{\partial T}$. 
This property is different from that of 
the ordinary second-order chiral phase 
transition, shown by the dotted curve of 
Fig. \ref{PhaseDiagram-RI}, that is discontinuous 
in both $ {\partial o_\alpha}/{\partial \theta} $ and 
$ {\partial o_\alpha}/{\partial T} $. 
The coexistence of the zeroth-order discontinuity of $n$ 
and the first-order one of $\sigma$ is originated from the fact 
that the RW transition line is vertical in the $\theta$-$T$ plane and 
does not move in the $\theta$ direction by changing the 
external parameter $\lambda_\alpha=m_0$. 
The influence of the RW discontinuity of $n$ 
to the Polyakov loop is discussed in the following.

In the imaginary $\mu$ region, physical quantities have 
a periodicity of $2\pi/3$ in $\theta$, 
when these are 
invariant under the extended ${\mathbb Z}_3$ transformation~\cite{Sakai}, 
\begin{align}
&e^{\pm i \theta} \to e^{\pm i \theta} e^{\pm i{2\pi k\over{3}}},\quad  
\Phi(\theta)  \to \Phi(\theta) e^{-i{2\pi k\over{3}}}, 
\notag\\
&\bar {\Phi}(\theta) \to \bar {\Phi}(\theta) e^{i{2\pi k\over{3}}} .
\label{eq:K2}
\end{align}
This is called the RW periodicity \cite{RW}. 
The thermodynamical potential $\Omega_{\rm PNJL}$ and 
$\sigma$, $s$ and $n$ are invariant 
under the extended ${\mathbb Z}_3$ symmetry, but 
$\Phi$ and $\bar {\Phi}$ are not \cite{Sakai}. 
However, this can be cured by introducing the modified Polyakov loop 
$\Psi=\Phi \exp(i\theta)$ 
invariant under the extended ${\mathbb Z}_3$ transformation. 
We then consider a period $0 \le \theta \le 2\pi/3$ without loss of 
generality.

\begin{figure}[htbp]
\begin{center}
 \includegraphics[width=0.4\textwidth]{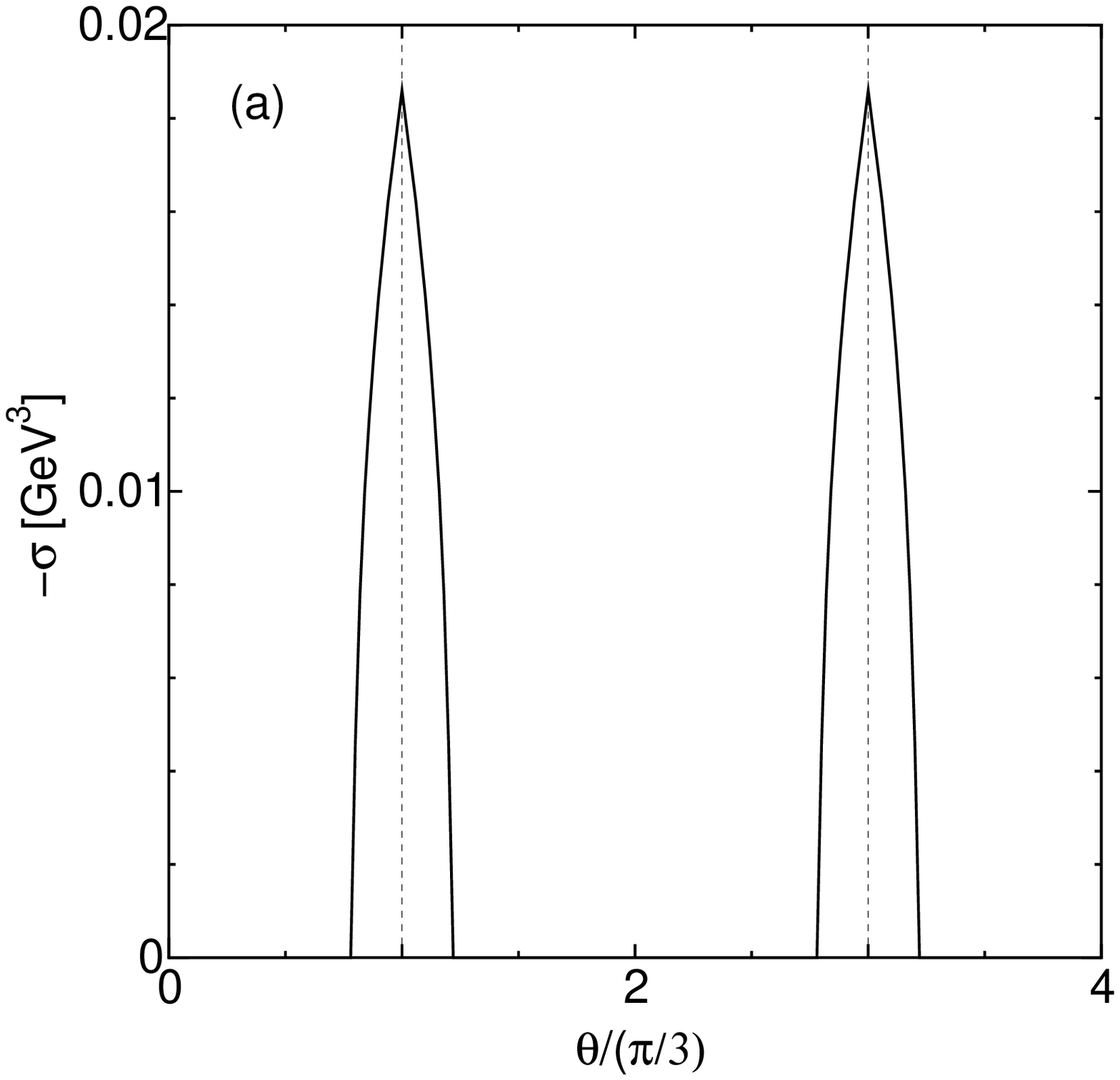} 
 \includegraphics[width=0.4\textwidth]{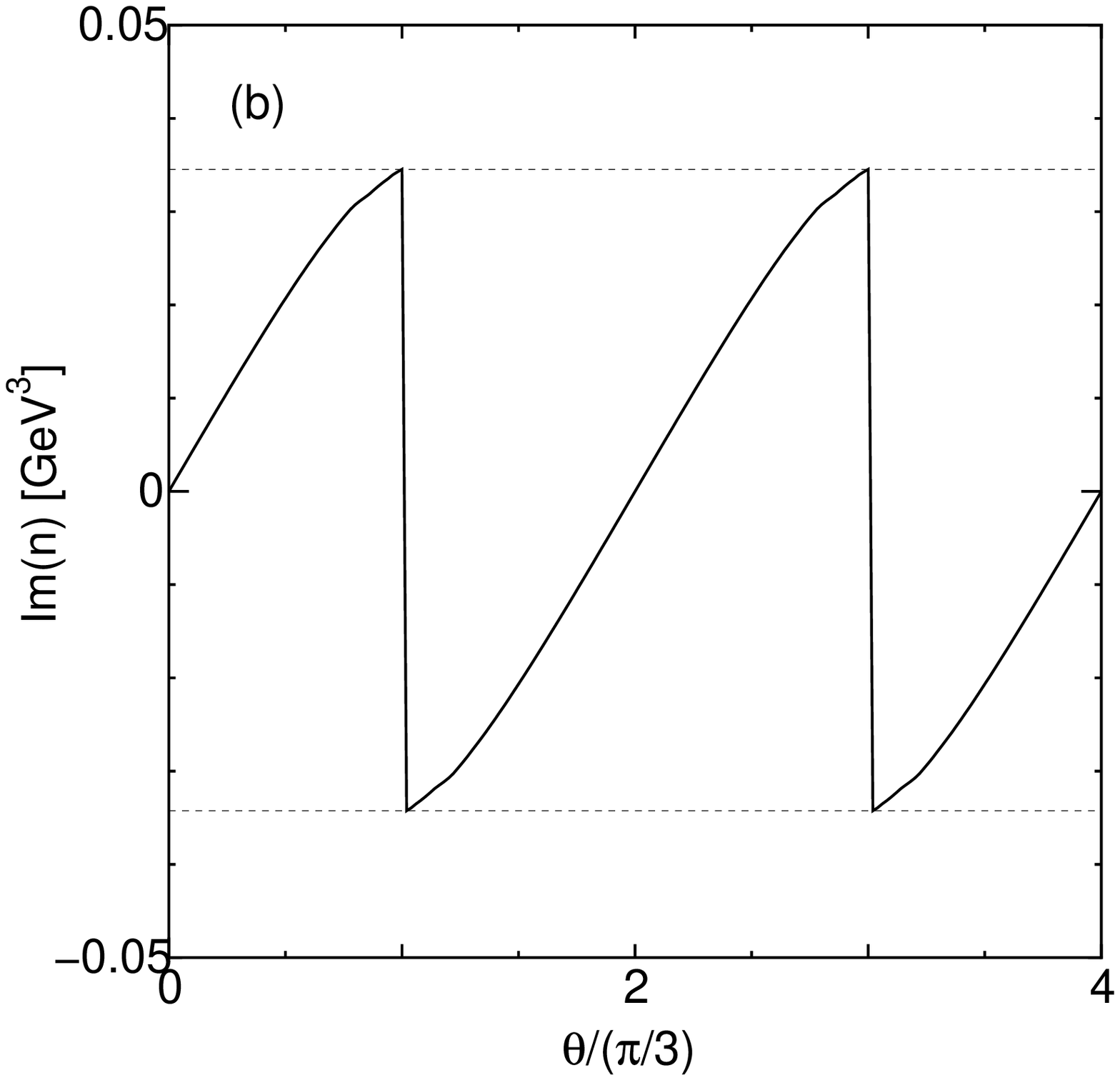} 
\end{center}
\caption{The $\theta$ dependence of the chiral condensate and 
the imaginary part of the quark number density 
at $T=300$~MeV in the chiral limit.}
\label{C-QD-I}
\end{figure}

Figure~\ref{C-QD-I} shows $\theta$ dependence of 
the chiral condensate $\sigma$ and 
the imaginary part of $n$, Im$[n]$, at $T=300$~MeV; 
note that $n$ is pure imaginary for imaginary $\mu$ by definition.
The chiral condensate has a cusp at 
$\theta=\pi/3$, while $n$ is discontinuous there. 
Thus, the first-order discontinuity of $\sigma$ and 
the zeroth-order discontinuity of $n$ coexist, as predicted above.

\begin{figure}[htbp]
\begin{center}
\includegraphics[width=0.4\textwidth]{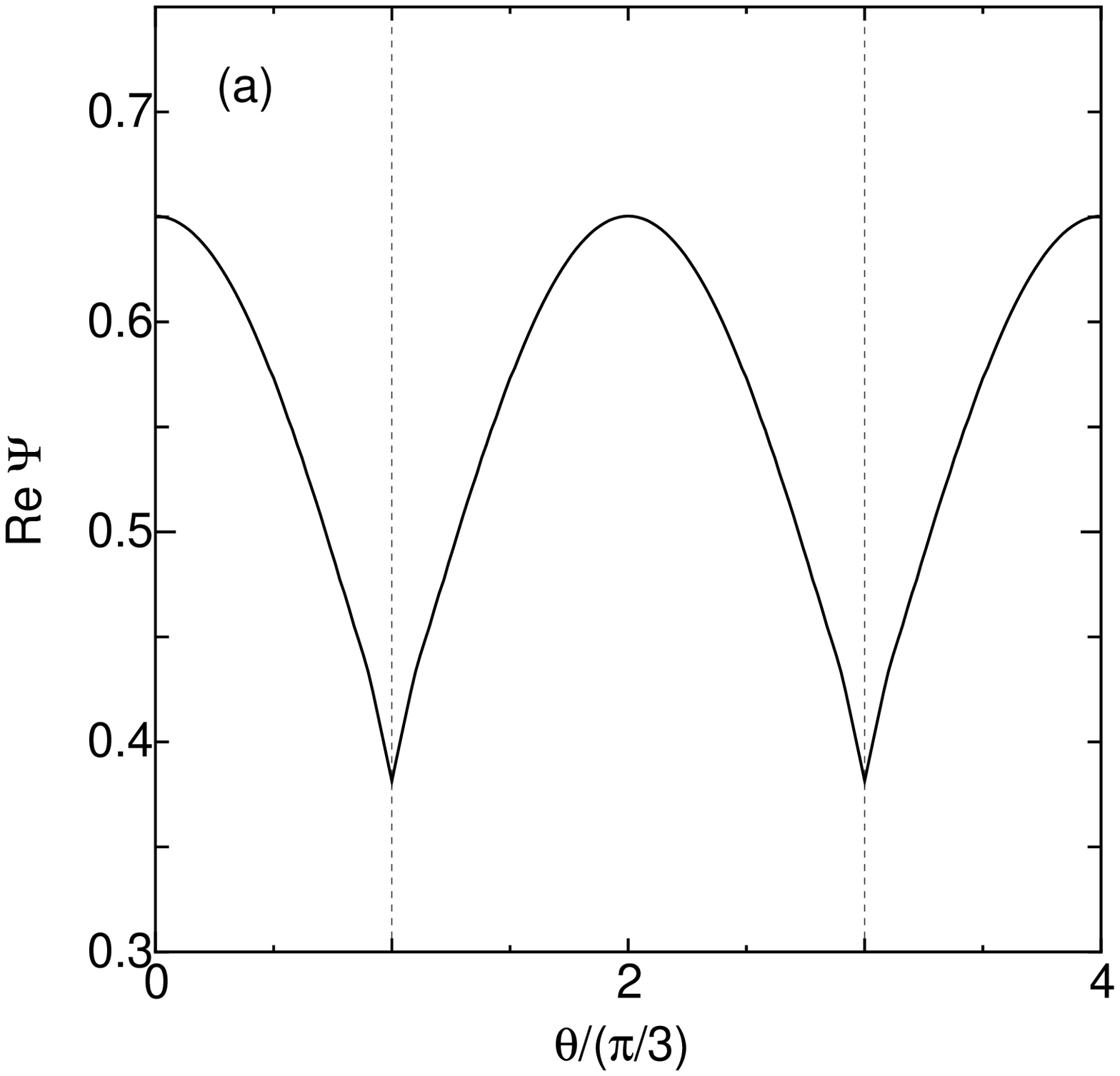}  
 \includegraphics[width=0.4\textwidth]{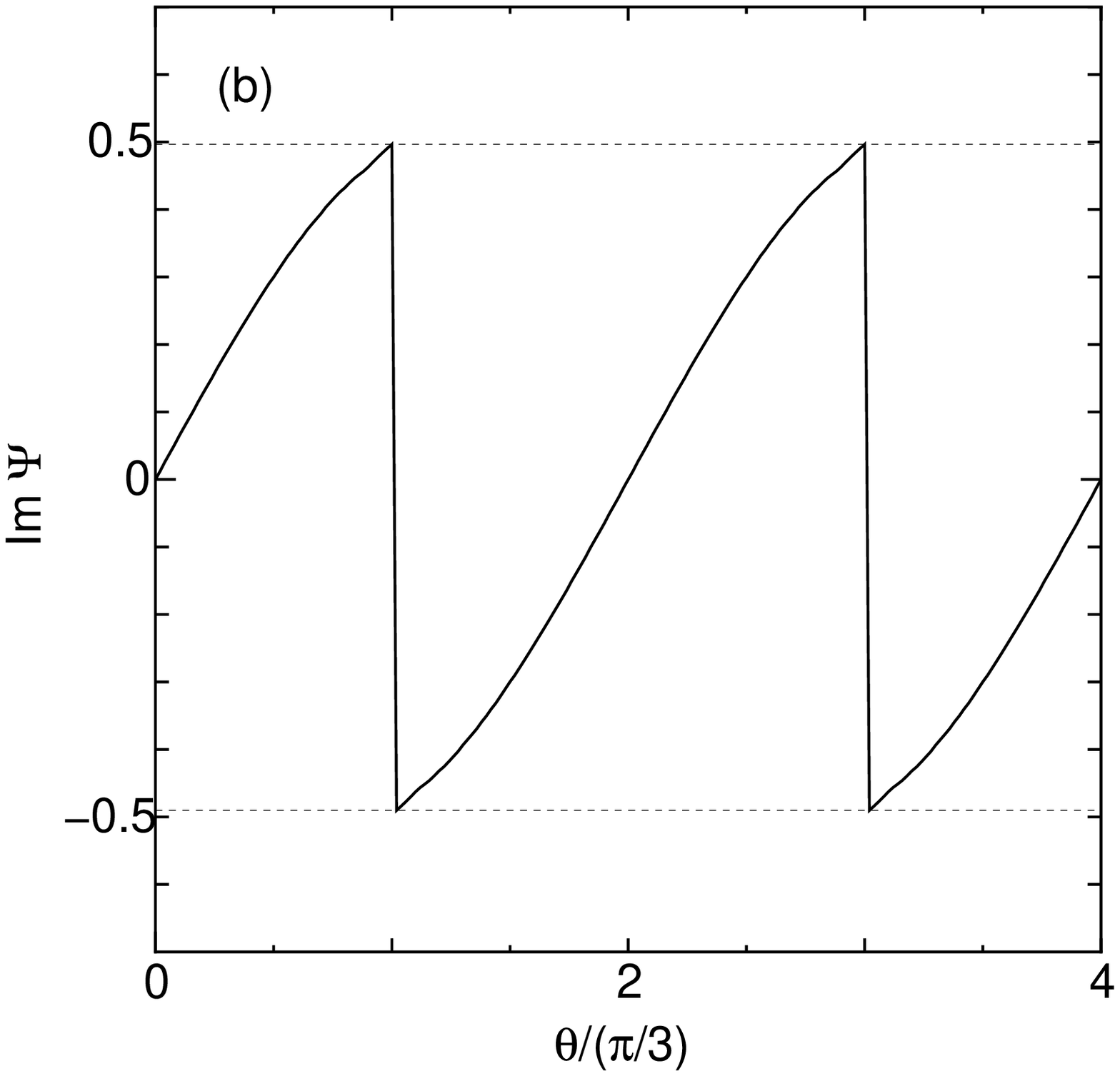}
\end{center}
\caption{The $\theta$ dependence of $\Psi$ at 
$T=300$~MeV in the chiral limit:(a) the real part and (b) the imaginary part.}
\label{CC-QD-I}
\end{figure}

Next, we consider the relation between 
the discontinuity of 
$n$ and the Polyakov loop transition by using 
${\rm Re}[\Psi]=(\Psi(\theta)+{\bar \Psi}(\theta))/2$. 
In the PNJL model, $\Psi(\theta)$ and $\bar{\Psi}(\theta)$ are 
treated as classical variables, 
and it is found from the expression for $\Omega_\mathrm{PNJL}$ 
in (13) of Ref.~\cite{Kashiwa1} that 
$\bar{\Psi}(\theta)$ is the complex conjugate of $\Psi(\theta)$ 
for the case of imaginary $\mu$. 
Figure~\ref{CC-QD-I}(a) shows $\theta$ dependence of 
the real part ${\rm Re}[\Psi]$ 
at $T=300$~MeV. There appears a first-order discontinuity also 
in ${\rm Re}[\Psi]$ on the line $\theta=\pi/3$, as expected from 
(\ref{1st-2nd-propagation-1}).

Finally, we consider the imaginary part of $\Psi$, 
${\rm Im}[\Psi]=(\Psi(\theta)-{\bar \Psi}(\theta))/2i$. 
This is also real, but $\theta$-odd (odd under the interchange of 
$\theta \leftrightarrow - \theta$ ), 
because $\Psi(\theta)=\bar{\Psi}(-\theta)$ \cite{Sakai}. 
One can not use $\lambda_\alpha {\rm Im}[\Psi]$ 
as a source term $\lambda_\alpha {\cal O}_\alpha$, 
since it breaks $\theta$-evenness, 
$\Omega_{\rm PNJL}(\theta)=\Omega_{\rm PNJL}(-\theta)$, that is 
the charge-conjugation symmetry of 
$\Omega_{\rm PNJL}$ \cite{Kratochvila}. 
To avoid this problem, 
we introduce a source term, $\lambda_\alpha \sin(3\theta) {\rm Im}[\Psi]$, 
designed to keep $\theta$-evenness and the RW periodicity. 
This is just an example of operators having the two properties. 
For this source term, 
(\ref{1st-2nd-propagation-1}) is reduced to 
\bea
\delta \Bigl( {\rm Im}[\Psi] \Bigr) 
= - \frac{1}{3} \frac{\partial f}{\partial \lambda_\alpha} \Bigl|_{\tilde{\rm c}} .
\label{coexistence-2}
\eea
Thus, $\delta ({\rm Im}[\Psi])$ is finite on 
the RW phase transition line $\theta=\pi/3$ because of
${\partial f}/{\partial \lambda_\alpha}|_{\tilde{\rm c}} \neq 0$. 
This indicates that the zeroth-order discontinuity of $n$ induces 
that of ${\rm Im}[\Psi]$ as shown in Fig~\ref{CC-QD-I}(b).

%
Throughout all the analyses, we can conclude that 
the zeroth-order discontinuity of a $\theta$-odd quantity $n$ 
induces zeroth-order ones in $\theta$-odd 
quantities and simultaneously does first-order ones 
in $\theta$-even quantities; 
see Ref.~\cite{Sakai} for the proof of the even/odd property of 
$n$, $\sigma$, ${\rm Re}[\Psi]$, ${\rm Im}[\Psi]$, $|\Psi|$ and arg$[\Psi]$. 
As shown in Fig. \ref{C-QD-I}, $\partial \sigma/\partial \theta$ is 
finite on both the sides of the critical chemical potential 
$\theta_{\rm c}$. This means that the chiral susceptibility $\chi$ 
is finite. Hence, there is no second-order phase transition on the RW line. 
Therefore, the RW phase transition is a first-order phase transition and 
a family of zeroth- and first-order discontinuities.

Figure~\ref{PhaseDiagram-I} shows the phase diagram on the $\theta$-$T$ plane 
that corresponds to the $\mu^2<0$ part of Fig. \ref{PhaseDiagram-RI}. 
On the dashed line between points A and B, 
the RW phase transition mentioned above emerges. 
The transition comes out also on the dashed line going up from 
point B,  although $\sigma$ is zero there and then 
no discontinuity takes place in $\sigma$. 
The dotted curves represents ordinary chiral phase transitions of second order. 

To summarize, we showed that once a zeroth- or first-order discontinuity 
takes place in a quantity $o_\gamma$, 
discontinuities of the same order emerge in other quantities 
$o_{\alpha \neq \gamma}$, 
if the conditions (\ref{condition-T}) and 
(\ref{condition-mu}) are satisfied, that is, 
if the phase boundary is shifted in both the directions of $T$ and $\mu$ 
in the $T$-$\mu$ plane by varying values of external parameters $\lambda_{\alpha}$ 
conjugate to $o_\alpha$. 
This coexistence theorem  is 
an extension of the BCPG theorem on 
the zeroth-order discontinuity of order parameter 
(the first-order phase transition). 
When the conditions break, first- and 
second-order discontinuities can coexist on the same phase boundary. 
The RW phase transition in the $\theta$-$T$ plane, composed of 
zeroth-order discontinuities of $\theta$-odd quantities and 
first-order discontinuities of $\theta$-even ones, 
is a typical example of the coexistence of 
zeroth- and first-order discontinuities. 
The RW phase transition line is vertical and 
does not move in the $\theta$ direction, 
even if any external parameter varies. 
Thus, the shape of the phase boundary and its variation 
with external parameters are essential in 
determining which type of coexistence is realized.

This work has been supported in part by the Grants-in-Aid for 
Scientific Research (18540280) of Education, Science, Sports, 
and Culture of Japan.



\begin{thebibliography}{19}
\expandafter\ifx\csname natexlab\endcsname\relax\def\natexlab#1{#1}\fi
\expandafter\ifx\csname bibnamefont\endcsname\relax
  \def\bibnamefont#1{#1}\fi
\expandafter\ifx\csname bibfnamefont\endcsname\relax
  \def\bibfnamefont#1{#1}\fi
\expandafter\ifx\csname citenamefont\endcsname\relax
  \def\citenamefont#1{#1}\fi
\expandafter\ifx\csname url\endcsname\relax
  \def\url#1{\texttt{#1}}\fi
\expandafter\ifx\csname urlprefix\endcsname\relax\def\urlprefix{URL }\fi
\providecommand{\bibinfo}[2]{#2}
\providecommand{\eprint}[2][]{\url{#2}}

%

\bibitem[{\citenamefont{Alford et al.}(2007)}]{Alford}
\bibinfo{author}{\bibfnamefont{M.}~\bibnamefont{G.}~\bibnamefont{Alford}},
\bibinfo{author}{\bibfnamefont{K.}~\bibnamefont{Rajagopal}},
\bibinfo{author}{\bibfnamefont{T.}~\bibnamefont{Schaefer}},
\bibnamefont{and} 
\bibinfo{author}{\bibfnamefont{A.}~\bibnamefont{Schmitt}}, 
\bibinfo{journal}{Rev. Mod. Phys.} \textbf{\bibinfo{volume}{80}},
\bibinfo{pages}{1455} (\bibinfo{year}{2008}).
%
\bibitem[{\citenamefont{Barducci et al.}(1993)}]{BCGG}
\bibinfo{author}{\bibfnamefont{A.}~\bibnamefont{Barducci}}, 
\bibinfo{author}{\bibfnamefont{R.}~\bibnamefont{Casalbuoni}}, 
\bibinfo{author}{\bibfnamefont{G.}~\bibnamefont{Pettini}}, 
\bibnamefont{and}
\bibinfo{author}{\bibfnamefont{R.}~\bibnamefont{Gatto}},  
\bibinfo{journal}{Phys. Lett.\ B} \textbf{\bibinfo{volume}{301}},
\bibinfo{pages}{95} (\bibinfo{year}{1993}). 

%
\bibitem[{\citenamefont{Kogut}(2007)}]{Kogut}
\bibinfo{author}{\bibfnamefont{J.}~\bibnamefont{B.}~\bibnamefont{Kogut}}, 
\bibnamefont{and} 
\bibinfo{author}{\bibfnamefont{D.}~\bibnamefont{K.}~\bibnamefont{Sinclair}},  
\bibinfo{journal}{Phys. Rev. D} \textbf{\bibinfo{volume}{77}},
\bibinfo{pages}{114503} (\bibinfo{year}{2008}).

%
\bibitem[{\citenamefont{Meisinger et al.}(1996)}]{Meisinger}
\bibinfo{author}{\bibfnamefont{P.}~\bibnamefont{N.}}~\bibnamefont{Meisinger},
\bibnamefont{and}
\bibinfo{author}{\bibfnamefont{M.}~\bibnamefont{C.}}~\bibnamefont{Ogilvie},  
  \bibinfo{journal}{Phys. Lett.\ B} \textbf{\bibinfo{volume}{379}},
  \bibinfo{pages}{163} (\bibinfo{year}{1996}). 

\bibitem[{\citenamefont{Dumitru}(2002)}]{Dumitru}
\bibinfo{author}{\bibfnamefont{A.}~\bibnamefont{Dumitru}},
\bibnamefont{and}
\bibinfo{author}{\bibfnamefont{R.}~\bibfnamefont{D.}~\bibnamefont{Pisarski}},  
\bibinfo{journal}{Phys.\ Rev.\  D} \textbf{\bibinfo{volume}{66}},
\bibinfo{pages}{096003} (\bibinfo{year}{2002}); 
\bibinfo{author}{\bibfnamefont{A.}~\bibnamefont{Dumitru}},
\bibinfo{author}{\bibfnamefont{Y.}~\bibnamefont{Hatta}},
\bibinfo{author}{\bibfnamefont{J.}~\bibnamefont{Lenaghan}},
\bibinfo{author}{\bibfnamefont{K.}~\bibnamefont{Orginos}},
\bibnamefont{and}
\bibinfo{author}{\bibfnamefont{R.}~\bibfnamefont{D.}~\bibnamefont{Pisarski}},  
\bibinfo{journal}{Phys.\ Rev.\  D} \textbf{\bibinfo{volume}{70}},
\bibinfo{pages}{034511} (\bibinfo{year}{2004}); 
\bibinfo{author}{\bibfnamefont{A.}~\bibnamefont{Dumitru}},
\bibinfo{author}{\bibfnamefont{R.}~\bibfnamefont{D.}~\bibnamefont{Pisarski}},  
\bibnamefont{and}
\bibinfo{author}{\bibfnamefont{D.}~\bibnamefont{Zschiesche}},  
\bibinfo{journal}{Phys.\ Rev.\  D} \textbf{\bibinfo{volume}{72}},
\bibinfo{pages}{065008} (\bibinfo{year}{2005}).

\bibitem[{\citenamefont{Fukushima}(2004)}]{Fukushima1}
\bibinfo{author}{\bibfnamefont{K.}~\bibnamefont{Fukushima}}, 
  \bibinfo{journal}{Phys. Lett.\ B} \textbf{\bibinfo{volume}{591}},
  \bibinfo{pages}{277} (\bibinfo{year}{2004});
  \bibinfo{journal}{Phys. Rev. D} \textbf{\bibinfo{volume}{78}},
  \bibinfo{pages}{114019} (\bibinfo{year}{2008}).

\bibitem[{\citenamefont{Fukushima}(2004)}]{Fukushima2}
\bibinfo{author}{\bibfnamefont{K.}~\bibnamefont{Fukushima}}, 
  \bibinfo{journal}{Phys. Rev. D} \textbf{\bibinfo{volume}{77}},
  \bibinfo{pages}{114028} (\bibinfo{year}{2008}).

\bibitem[{\citenamefont{{S. K. Ghosh} et al.}(2006)}]{Ghos}
\bibinfo{author}{\bibnamefont{{S. K. Ghosh}}},
  \bibinfo{author}{\bibnamefont{{T. K. Mukherjee}}},
  \bibinfo{author}{\bibnamefont{{M. G. Mustafa}}}, \bibnamefont{and}
  \bibinfo{author}{\bibfnamefont{R.}~\bibnamefont{Ray}},
  \bibinfo{journal}{Phys.\ Rev.\ D} \textbf{\bibinfo{volume}{73}},
  \bibinfo{pages}{114007} (\bibinfo{year}{2006}). 

\bibitem[{\citenamefont{Megias et al.}(2006)}]{Megias}
\bibinfo{author}{\bibfnamefont{E.}~\bibnamefont{Meg{$\acute{\i}$}as}},
\bibinfo{author}{\bibfnamefont{E.}~\bibnamefont{R.}~\bibnamefont{Arriola}},
\bibnamefont{and}
\bibinfo{author}{\bibfnamefont{L.}~\bibnamefont{L.}~\bibnamefont{Salcedo}},  
  \bibinfo{journal}{Phys. Rev.\ D} \textbf{\bibinfo{volume}{74}},
  \bibinfo{pages}{065005} (\bibinfo{year}{2006}). 

\bibitem[{\citenamefont{Ratti et al.}(2006)}]{Ratti1}
\bibinfo{author}{\bibfnamefont{C.}~\bibnamefont{Ratti}},
\bibinfo{author}{\bibfnamefont{M.}~\bibfnamefont{A.}~\bibnamefont{Thaler}},
\bibnamefont{and}
\bibinfo{author}{\bibfnamefont{W.}~\bibnamefont{Weise}},  
  \bibinfo{journal}{Phys. Rev.\ D} \textbf{\bibinfo{volume}{73}},
  \bibinfo{pages}{014019} (\bibinfo{year}{2006}). 

\bibitem[{\citenamefont{Ciminale}(2007)}]{Ciminale}
\bibinfo{author}{\bibfnamefont{M.}~\bibnamefont{Ciminale}},
\bibinfo{author}{\bibfnamefont{R.}~\bibnamefont{Gatto}},
\bibinfo{author}{\bibfnamefont{G.}~\bibnamefont{Nardulli}},
\bibnamefont{and}
\bibinfo{author}{\bibfnamefont{M.}~\bibnamefont{Ruggieri}},
  \bibinfo{journal}{Phys.\ Lett.\ B} \textbf{\bibinfo{volume}{657}},
  \bibinfo{pages}{64} (\bibinfo{year}{2007});
\bibinfo{author}{\bibfnamefont{M.}~\bibnamefont{Ciminale}},
\bibinfo{author}{\bibfnamefont{R.}~\bibnamefont{Gatto}},
\bibinfo{author}{\bibfnamefont{N.}~\bibfnamefont{D.}~\bibnamefont{Ippolito}},
\bibinfo{author}{\bibfnamefont{G.}~\bibnamefont{Nardulli}},  
\bibnamefont{and}
\bibinfo{author}{\bibfnamefont{M.}~\bibnamefont{Ruggieri}},  
  \bibinfo{journal}{Phys. Rev.\ D} \textbf{\bibinfo{volume}{77}},
  \bibinfo{pages}{054023} (\bibinfo{year}{2008}).

\bibitem[{\citenamefont{Ratti et al.}(2007)}]{Ratti2}
\bibinfo{author}{\bibfnamefont{C.}~\bibnamefont{Ratti}},
\bibinfo{author}{\bibfnamefont{S.}~\bibnamefont{R\"{o}{\ss}ner}},
\bibinfo{author}{\bibfnamefont{M.}~\bibfnamefont{A.}~\bibnamefont{Thaler}},
\bibnamefont{and}
\bibinfo{author}{\bibfnamefont{W.}~\bibnamefont{Weise}},  
  \bibinfo{journal}{Eur. Phys. J.\ C} \textbf{\bibinfo{volume}{49}},
  \bibinfo{pages}{213} (\bibinfo{year}{2007}). 

\bibitem[{\citenamefont{Rossner et al.}(2007)}]{Rossner}
\bibinfo{author}{\bibfnamefont{S.}~\bibnamefont{R\"{o}{\ss}ner}},
\bibinfo{author}{\bibfnamefont{C.}~\bibnamefont{Ratti}},
\bibnamefont{and}
\bibinfo{author}{\bibfnamefont{W.}~\bibnamefont{Weise}},  
  \bibinfo{journal}{Phys. Rev.\ D} \textbf{\bibinfo{volume}{75}},
  \bibinfo{pages}{034007} (\bibinfo{year}{2007});
\bibinfo{author}{\bibfnamefont{S.}~\bibnamefont{R\"{o}{\ss}ner}},
\bibinfo{author}{\bibfnamefont{T.}~\bibnamefont{Hell}},
\bibinfo{author}{\bibfnamefont{C.}~\bibnamefont{Ratti}},
\bibnamefont{and}
\bibinfo{author}{\bibfnamefont{W.}~\bibnamefont{Weise}},
  \bibinfo{journal}{Nucl. Phys.} \textbf{\bibinfo{volume}{A814}},
  \bibinfo{pages}{118} (\bibinfo{year}{2008}).

\bibitem[{\citenamefont{Hansen et al.}(2007)}]{Hansen}
\bibinfo{author}{\bibfnamefont{H.}~\bibnamefont{Hansen}}, 
\bibinfo{author}{\bibfnamefont{W.}~\bibfnamefont{M.}~\bibnamefont{Alberico}},
\bibinfo{author}{\bibfnamefont{A.}~\bibnamefont{Beraudo}}, 
\bibinfo{author}{\bibfnamefont{A.}~\bibnamefont{Molinari}},
\bibinfo{author}{\bibfnamefont{M.}~\bibnamefont{Nardi}},
\bibnamefont{and}
\bibinfo{author}{\bibfnamefont{C.}~\bibnamefont{Ratti}}, 
  \bibinfo{journal}{Phys. Rev.\ D} \textbf{\bibinfo{volume}{75}},
  \bibinfo{pages}{065004} (\bibinfo{year}{2007}). 

\bibitem[{\citenamefont{Sasaki et al.}(2007)}]{Sasaki}
\bibinfo{author}{\bibfnamefont{C.}~\bibnamefont{Sasaki}},
\bibinfo{author}{\bibfnamefont{B.}~\bibnamefont{Friman}},
\bibnamefont{and}
\bibinfo{author}{\bibfnamefont{K.}~\bibnamefont{Redlich}}, 
\bibinfo{journal}{Phys. Rev.\ D} \textbf{\bibinfo{volume}{75}},
  \bibinfo{pages}{074013} (\bibinfo{year}{2007}). 

\bibitem[{\citenamefont{Schaefer}(2007)}]{Schaefer}
\bibinfo{author}{\bibfnamefont{B.}~\bibfnamefont{-J.}~\bibnamefont{Schaefer}},
\bibinfo{author}{\bibfnamefont{J.}~\bibfnamefont{M.}~\bibnamefont{Pawlowski}},
\bibnamefont{and}
\bibinfo{author}{\bibfnamefont{J.}~\bibnamefont{Wambach}},  
  \bibinfo{journal}{Phys.\ Rev.\  D} \textbf{\bibinfo{volume}{76}},
  \bibinfo{pages}{074023} (\bibinfo{year}{2007}).


\bibitem[{\citenamefont{Costa et al}(2009)}]{Costa}
\bibinfo{author}{\bibfnamefont{P.}~\bibnamefont{Costa}}, 
\bibinfo{author}{\bibfnamefont{C.}~\bibfnamefont{A.}~\bibfnamefont{de}~\bibnamefont{Sousa}}, 
\bibinfo{author}{\bibfnamefont{M.}~\bibfnamefont{C.}~\bibnamefont{Ruivo}}, 
\bibnamefont{and}
\bibinfo{author}{\bibfnamefont{H.}~\bibnamefont{Hansen}},
  \bibinfo{journal}{Europhys. Lett.} \textbf{\bibinfo{volume}{86}},
  \bibinfo{pages}{31001} (\bibinfo{year}{2009});
\bibinfo{author}{\bibfnamefont{P.}~\bibnamefont{Costa}}, 
\bibinfo{author}{\bibfnamefont{M.}~\bibfnamefont{C.}~\bibnamefont{Ruivo}}, 
\bibinfo{author}{\bibfnamefont{C.}~\bibfnamefont{A.}~\bibfnamefont{de}~\bibnamefont{Sousa}}, 
\bibinfo{author}{\bibfnamefont{H.}~\bibnamefont{Hansen}},
\bibnamefont{and}
\bibinfo{author}{\bibfnamefont{W.}~\bibfnamefont{M.}~\bibnamefont{Alberico}},
  \bibinfo{journal}{Phys.\ Rev.\  D} \textbf{\bibinfo{volume}{79}},
  \bibinfo{pages}{116003} (\bibinfo{year}{2009}).

\bibitem[{\citenamefont{Kashiwa et al}(2008)}]{Kashiwa1}
\bibinfo{author}{\bibfnamefont{K.}~\bibnamefont{Kashiwa}}, 
\bibinfo{author}{\bibfnamefont{H.}~\bibnamefont{Kouno}}, 
\bibinfo{author}{\bibfnamefont{M.}~\bibnamefont{Matsuzaki}}, 
\bibnamefont{and}
\bibinfo{author}{\bibfnamefont{M.}~\bibnamefont{Yahiro}},
  \bibinfo{journal}{Phys.\ Lett.\ B} \textbf{\bibinfo{volume}{662}},
  \bibinfo{pages}{26} (\bibinfo{year}{2008});
%
\bibinfo{author}{\bibfnamefont{K.}~\bibnamefont{Kashiwa}},
\bibinfo{author}{\bibfnamefont{M.}~\bibnamefont{Matsuzaki}}, 
\bibinfo{author}{\bibfnamefont{H.}~\bibnamefont{Kouno}}, 
\bibinfo{author}{\bibfnamefont{Y.}~\bibnamefont{Sakai}},  
\bibnamefont{and}
\bibinfo{author}{\bibfnamefont{M.}~\bibnamefont{Yahiro}},
  \bibinfo{journal}{Phys.\ Rev.\  D} \textbf{\bibinfo{volume}{79}},
  \bibinfo{pages}{076008} (\bibinfo{year}{2009}).

\bibitem[{\citenamefont{Fu}(2007)}]{Fu}
\bibinfo{author}{\bibfnamefont{W.}~\bibfnamefont{J.}~\bibnamefont{Fu}},
\bibinfo{author}{\bibfnamefont{Z.}~\bibnamefont{Zhang}},
\bibnamefont{and}
\bibinfo{author}{\bibfnamefont{Y.}~\bibfnamefont{X.}~\bibnamefont{Liu}},
  \bibinfo{journal}{Phys.\ Rev.\  D} \textbf{\bibinfo{volume}{77}},
  \bibinfo{pages}{014006} (\bibinfo{year}{2008}).

\bibitem[{\citenamefont{Abuki}(2008)}]{Abuki1}
\bibinfo{author}{\bibfnamefont{H.}~\bibnamefont{Abuki}},
\bibinfo{author}{\bibfnamefont{M.}~\bibnamefont{Ciminale}},
\bibinfo{author}{\bibfnamefont{R.}~\bibnamefont{Gatto}},
\bibinfo{author}{\bibfnamefont{G.}~\bibnamefont{Nardulli}},
\bibnamefont{and}
\bibinfo{author}{\bibfnamefont{M.}~\bibnamefont{Ruggieri}},
 \bibinfo{journal}{Phys.\ Rev.\  D} \textbf{\bibinfo{volume}{77}},
  \bibinfo{pages}{074018} (\bibinfo{year}{2008});
%
\bibinfo{author}{\bibfnamefont{H.}~\bibnamefont{Abuki}},
\bibinfo{author}{\bibfnamefont{M.}~\bibnamefont{Ciminale}},
\bibinfo{author}{\bibfnamefont{R.}~\bibnamefont{Gatto}},
\bibinfo{author}{\bibfnamefont{N.}~\bibfnamefont{D.}~\bibnamefont{Ippolito}},
\bibinfo{author}{\bibfnamefont{G.}~\bibnamefont{Nardulli}},
\bibnamefont{and}
\bibinfo{author}{\bibfnamefont{M.}~\bibnamefont{Ruggieri}},
 \bibinfo{journal}{Phys.\ Rev.\  D} \textbf{\bibinfo{volume}{78}},
  \bibinfo{pages}{014002} (\bibinfo{year}{2008});  
%
\bibinfo{author}{\bibfnamefont{H.}~\bibnamefont{Abuki}},
\bibinfo{author}{\bibfnamefont{M.}~\bibnamefont{Ciminale}},
\bibinfo{author}{\bibfnamefont{R.}~\bibnamefont{Gatto}},
\bibnamefont{and}
\bibinfo{author}{\bibfnamefont{M.}~\bibnamefont{Ruggieri}}, 
  \bibinfo{journal}{Phys.\ Rev.\  D} \textbf{\bibinfo{volume}{79}},
  \bibinfo{pages}{034021} (\bibinfo{year}{2009}).


\bibitem[{\citenamefont{Abuki}(2008)}]{Abuki2}
\bibinfo{author}{\bibfnamefont{H.}~\bibnamefont{Abuki}},
\bibinfo{author}{\bibfnamefont{R.}~\bibnamefont{Anglani}},
\bibinfo{author}{\bibfnamefont{R.}~\bibnamefont{Gatto}},
\bibinfo{author}{\bibfnamefont{G.}~\bibnamefont{Nardulli}},
\bibnamefont{and}
\bibinfo{author}{\bibfnamefont{M.}~\bibnamefont{Ruggieri}},
 \bibinfo{journal}{Phys.\ Rev.\  D} \textbf{\bibinfo{volume}{78}},
  \bibinfo{pages}{034034} (\bibinfo{year}{2008}).


\bibitem[{\citenamefont{Sakai et al}(2008)}]{Sakai}
\bibinfo{author}{\bibfnamefont{Y.}~\bibnamefont{Sakai}},
\bibinfo{author}{\bibfnamefont{K.}~\bibnamefont{Kashiwa}}, 
\bibinfo{author}{\bibfnamefont{H.}~\bibnamefont{Kouno}}, 
\bibnamefont{and}
\bibinfo{author}{\bibfnamefont{M.}~\bibnamefont{Yahiro}},
  \bibinfo{journal}{Phys.\ Rev.\  D} \textbf{\bibinfo{volume}{77}},
  \bibinfo{pages}{051901(R)} (\bibinfo{year}{2008});
\bibinfo{journal}{Phys.\ Rev.\  D} \textbf{\bibinfo{volume}{78}},
  \bibinfo{pages}{036001} (\bibinfo{year}{2008});
\bibinfo{author}{\bibfnamefont{Y.}~\bibnamefont{Sakai}},
\bibinfo{author}{\bibfnamefont{K.}~\bibnamefont{Kashiwa}}, 
\bibinfo{author}{\bibfnamefont{H.}~\bibnamefont{Kouno}},
\bibinfo{author}{\bibfnamefont{M.}~\bibnamefont{Matsuzaki}},
\bibnamefont{and}
\bibinfo{author}{\bibfnamefont{M.}~\bibnamefont{Yahiro}},
  \bibinfo{journal}{Phys.\ Rev.\  D} \textbf{\bibinfo{volume}{78}},
  \bibinfo{pages}{076007} (\bibinfo{year}{2008});
\bibinfo{journal}{Phys.\ Rev.\  D} \textbf{\bibinfo{volume}{79}},
  \bibinfo{pages}{096001} (\bibinfo{year}{2009}). 


\bibitem[{\citenamefont{Kouno et al}(2009)}]{Kouno}
\bibinfo{author}{\bibfnamefont{H.}~\bibnamefont{Kouno}}, 
\bibinfo{author}{\bibfnamefont{Y.}~\bibnamefont{Sakai}},
\bibinfo{author}{\bibfnamefont{K.}~\bibnamefont{Kashiwa}}, 
\bibnamefont{and}
\bibinfo{author}{\bibfnamefont{M.}~\bibnamefont{Yahiro}},
 \bibinfo{howpublished}{arXiv:hep-ph/0904.0925 } (\bibinfo{year}{2009}). 


\bibitem[{\citenamefont{Fujii}(2006)}]{Fujii}
\bibinfo{author}{\bibfnamefont{H.}~\bibnamefont{Fujii}},  
\bibinfo{journal}{Phys. Rev.\ D} \textbf{\bibinfo{volume}{67}},
\bibinfo{pages}{094018} (\bibinfo{year}{2003}). 
%
\bibitem[{\citenamefont{Fujii and Ohtani}(2006)}]{Fujii2}
\bibinfo{author}{\bibfnamefont{H.}~\bibnamefont{Fujii}},
\bibnamefont{and}
\bibinfo{author}{\bibfnamefont{M.}~\bibnamefont{Ohtani}},  
\bibinfo{journal}{Phys. Rev.\ D} \textbf{\bibinfo{volume}{70}},
\bibinfo{pages}{014016} (\bibinfo{year}{2004}). 




\bibitem[{\citenamefont{Roberge and Weiss}(1986)}]{RW}
\bibinfo{author}{\bibfnamefont{A.}~\bibnamefont{Roberge}} \bibnamefont{and}
\bibinfo{author}{\bibfnamefont{N.}~\bibnamefont{Weiss}},  
\bibinfo{journal}{Nucl. Phys. } \textbf{\bibinfo{volume}{B275}},
\bibinfo{pages}{734} (\bibinfo{year}{1986}). 

\bibitem[{\citenamefont{Halasz et al.}(1993)}]{Halasz}
\bibinfo{author}{\bibfnamefont{M.}~\bibnamefont{A.}~\bibnamefont{Halasz}}, 
\bibinfo{author}{\bibfnamefont{A.}~\bibnamefont{D.}~\bibnamefont{Jackson}}, 
\bibinfo{author}{\bibfnamefont{R.}~\bibnamefont{E.}~\bibnamefont{Shrock}}, 
\bibinfo{author}{\bibfnamefont{M.}~\bibnamefont{A.}~\bibnamefont{Stephanov}}, 
\bibnamefont{and}
\bibinfo{author}{\bibfnamefont{J.}~\bibnamefont{J.}~\bibnamefont{M.}~\bibnamefont{Verbaarschot}}, 
\bibinfo{journal}{Phys.\ Rev.\  D} \textbf{\bibinfo{volume}{58}},
\bibinfo{pages}{096007} (\bibinfo{year}{1998}). 

\bibitem[{\citenamefont{Tawfik and Toublan}(2005)}]{Tawfik}
\bibinfo{author}{\bibfnamefont{A.}~\bibnamefont{Tawfik}} \bibnamefont{and}
\bibinfo{author}{\bibfnamefont{D.}~\bibnamefont{Toublan}},  
\bibinfo{journal}{Phys. Lett.\ B} \textbf{\bibinfo{volume}{623}},
\bibinfo{pages}{48} (\bibinfo{year}{2005}). 

\bibitem[{\citenamefont{Frank et al.}(2003)}]{Frank}
\bibinfo{author}{\bibfnamefont{M.}~\bibnamefont{Frank}},
\bibinfo{author}{\bibfnamefont{M.}~\bibnamefont{Buballa}} \bibnamefont{and}
\bibinfo{author}{\bibfnamefont{M.}~\bibnamefont{Oertel}},  
\bibinfo{journal}{Phys. Lett.\ B} \textbf{\bibinfo{volume}{562}},
\bibinfo{pages}{221} (\bibinfo{year}{2003}). 

\bibitem[{\citenamefont{Kashiwa et al}(2006)}]{Kashiwa2}
\bibinfo{author}{\bibfnamefont{K.}~\bibnamefont{Kashiwa}}, 
\bibinfo{author}{\bibfnamefont{H.}~\bibnamefont{Kouno}}, 
\bibinfo{author}{\bibfnamefont{T.}~\bibnamefont{Sakaguchi}}, 
\bibinfo{author}{\bibfnamefont{M.}~\bibnamefont{Matsuzaki}}, 
\bibnamefont{and}
\bibinfo{author}{\bibfnamefont{M.}~\bibnamefont{Yahiro}},
\bibinfo{journal}{Phys. Lett.\ B} \textbf{\bibinfo{volume}{647}},
\bibinfo{pages}{446} (\bibinfo{year}{2007}); 
\bibinfo{author}{\bibfnamefont{K.}~\bibnamefont{Kashiwa}}, 
\bibinfo{author}{\bibfnamefont{M.}~\bibnamefont{Matsuzaki}}, 
\bibinfo{author}{\bibfnamefont{H.}~\bibnamefont{Kouno}}, 
\bibnamefont{and}
\bibinfo{author}{\bibfnamefont{M.}~\bibnamefont{Yahiro}},
\bibinfo{journal}{Phys. Lett.\ B} \textbf{\bibinfo{volume}{657}},
\bibinfo{pages}{143} (\bibinfo{year}{2007}). 

\bibitem[{\citenamefont{Boyd et al.}(1996)}]{Boyd}
\bibinfo{author}{\bibfnamefont{G.}~\bibnamefont{Boyd}},
\bibinfo{author}{\bibfnamefont{J.}~\bibnamefont{Engels}},
\bibinfo{author}{\bibfnamefont{F.}~\bibnamefont{Karsch}},
\bibinfo{author}{\bibfnamefont{E.}~\bibnamefont{Laermann}},
\bibinfo{author}{\bibfnamefont{C.}~\bibnamefont{Legeland}},
\bibinfo{author}{\bibfnamefont{M.}~\bibnamefont{L\"{u}tgemeier}},
\bibnamefont{and}
\bibinfo{author}{\bibfnamefont{B.}~\bibnamefont{Petersson}},
 \bibinfo{journal}{Nucl. Phys.} \textbf{\bibinfo{volume}{B469}},
\bibinfo{pages}{419} (\bibinfo{year}{1996}). 

\bibitem[{\citenamefont{Kaczmarek}(2002)}]{Kaczmarek}
\bibinfo{author}{\bibfnamefont{O.}~\bibnamefont{Kaczmarek}},
\bibinfo{author}{\bibfnamefont{F.}~\bibnamefont{Karsch}},
\bibinfo{author}{\bibfnamefont{P.}~\bibnamefont{Petreczky}},
\bibnamefont{and}
\bibinfo{author}{\bibfnamefont{F.}~\bibnamefont{Zantow}},  
  \bibinfo{journal}{Phys. Lett.\ B} \textbf{\bibinfo{volume}{543}},
  \bibinfo{pages}{41} (\bibinfo{year}{2002}).


\bibitem[{\citenamefont{MacLerran}(2007)}]{McLerran1}
\bibinfo{author}{\bibfnamefont{L.}~\bibnamefont{McLerran}},
\bibnamefont{and}
\bibinfo{author}{\bibfnamefont{R.}~\bibfnamefont{D}~\bibnamefont{Pisarski}},  
  \bibinfo{journal}{Nucl. Phys.} \textbf{\bibinfo{volume}{A796}},
  \bibinfo{pages}{83} (\bibinfo{year}{2007}).


\bibitem[{\citenamefont{Miura}(2007)}]{Miura}
\bibinfo{author}{\bibfnamefont{K.}~\bibnamefont{Miura}},
\bibnamefont{and}
\bibinfo{author}{\bibfnamefont{A.}~\bibnamefont{Ohnishi}},  
  \bibinfo{howpublished}{arXiv:nucl-th/0806.3357} (\bibinfo{year}{2008}). 

\bibitem[{\citenamefont{MacLerran}(2008)}]{McLerran2}
\bibinfo{author}{\bibfnamefont{L.}~\bibnamefont{McLerran}},
\bibnamefont{and}
\bibinfo{author}{\bibfnamefont{K.}~\bibnamefont{Redlich}},  
\bibnamefont{and}
\bibinfo{author}{\bibfnamefont{C.}~\bibnamefont{Sasaki}},  
  \bibinfo{howpublished}{arXiv:hep-ph/0812.3585} (\bibinfo{year}{2008}). 


\bibitem[{\citenamefont{Kratochvila}(2006)}]{Kratochvila}
\bibinfo{author}{\bibfnamefont{S.}~\bibnamefont{Kratochvila}},
\bibnamefont{and}
\bibinfo{author}{\bibfnamefont{P.}~\bibfnamefont{de}~\bibnamefont{Forcrand}},  
  \bibinfo{journal}{Phys.\ Rev.\  D} \textbf{\bibinfo{volume}{73}},
  \bibinfo{pages}{114512} (\bibinfo{year}{2006}).



\end{thebibliography}
\end{document}